\documentclass[aps,nofootinbib,endfloats]{revtex4}
\usepackage{epsfig}
\input epsf.tex
\usepackage[english]{babel}
\usepackage{color}
\newcommand{\parl}{\parallel}
\usepackage{graphicx}
\usepackage{amsmath}

\begin{document}
\draft
\title{Magnetooptical Properties of Rydberg Excitons - Center-of-Mass Quantization Approach}
\author{Sylwia Zieli\'{n}ska-Raczy\'{n}ska, David Ziemkiewicz,
and Gerard Czajkowski } \affiliation{ Institute of Mathematics and
Physics, UTP University of Science and Technology, Al. Prof. S.
Kaliskiego 7, PL 85-789 Bydgoszcz
 (Poland)}
\begin{abstract}  We show how to compute the magnetooptical  functions (absorption, reflection, and transmission) when {Rydberg
Exciton-Polaritons} appear, including the effect of the coherence
between the electron-hole pair and the electromagnetic  field, and
the polaritonic effect. Using the Real Density Matrix Approach the
analytical expressions for magnetooptical functions are obtained and  numerical calculations
for Cu$_2$0 crystal are performed. The influence of the strength
of applied external magnetic field  on the resonance displacement
of excitonic spectra is discussed. We report a good agreement with
recently published experimental data.\end{abstract} \maketitle
\section{Introduction}
The concept of excitons has been formulated more than 80 years ago
by Y. Frenkel \cite{Frenkel}, who predicted their existence in
molecular crystals. A few years later, G. Wannier \cite{Wannier}
and N. de Mott \cite{Mott} described these electron-hole bound
states for inorganic semiconductors. In 1952 E. Gross and N.
Karriiev \cite{Gross} discovered such Wannier-Mott excitons
experimentally in a copper oxide semiconductor. After that time
excitons remain an important topic of experimental and theoretical
research, since they play a dominant role in the optical
properties of semiconductors (molecular crystals etc.,). The
excitons have been studied in great details in various types of
semiconductor nanostructures and in bulk crystals. Since there is
a large number of papers,  monographs, review articles devoted to
excitons, we refer to only a small collection of them
\cite{Knox}-\cite{Agran2009}. In the past decades the main effort
of the researchers was focussed on excitons in nanostructures
\cite{Weisbuch}-\cite{Harrison}, but very recently, new attention
has been drawn back to the subject of excitons in bulk crystals by
an experimental observation of the so-called yellow exciton series
in Cu$_2$O up to a large principal quantum number of \emph{n} = 25
\cite{Kazimierczuk}.  Such excitons in copper oxide, in analogy to
atomic physics, have been named Rydberg excitons. By virtue of
their special properties Rydberg excitons are of fascination in
solid and optical physics. These objects whose size scales as the
square of the Rydberg principal quantum number \emph{n}, are
ideally suited for fundamental quantum interrogations, as well as
detailed classical analysis.
 Several
theoretical approaches to calculate optical properties of Rydberg
excitons have been presented
\cite{Hofling}-\cite{Schweiner.Symmetry}. Recently quantum
coherence of Rydberg excitons in this system has been investigated
\cite{Gruenewald} which opens new avenue for their further
implementation in quantum information processing.

 When external constant
fields (electric or/and magnetic) are applied, the Rydberg
excitons, especially those with high principal number \emph{n},
show effects which are not observable in exciton systems with a
few number of excitonic states. This effects range from a large
Stark shift, overlapping of states, creation of higher order
excitons (F, H, etc) to quantum chaos and new type statistics for
exciton states \cite{assmann}. Although the Stark and other
electrooptic effects on Rydberg excitons in Cu$_2$O have been
measured and analized \cite{Thewes}, \cite{Heck},
\cite{Zielinska.PRB.2016.b}, there are only few results available
regarding the magnetooptic properties of Rydberg excitons
\cite{assmann} where
 the excitonic spectra of coprous oxide
subjected to an external magnetic field up to 7 T  has been measured and the
complex splitting pattern of crossing and overlapping levels has been demonstrated.

 Highly excited Rydberg excitons in Cu$_2$O crystal
provide a well-accessible venue for combined theoretical and
experimental studies of  magnetic field effects on the systems.
From a different perspective magnetic fields may offer a promising
possibility for a controlled manipulation of Rydberg excitons,
which would be otherwise difficult to trap by standard optical
techniques, developed for ground states. Due to the fact that free-space Rydberg polaritons have recently drawn intense
interest as tools for quantum information processing one can expected Rydberg excitons in solid may
become highly required object for creating high-fidelity photonic quantum materials \cite{jia}.
Magnetic fields can strongly affect the Rydberg-Rydberg interactions by breaking the Zeeman degeneracy that produces Foster zeros.

 In the presented paper we will focus  on
magnetooptical properties of Rydberg excitons in Cu$_2$O motivated
by the results presented in \cite{assmann}. As in our  previous
papers \cite{Zielinska.PRB}, \cite{Zielinska.PRB.2016.b}, we will
use the  method based on the Real Density Matrix Approach (RDMA).
Our  main purpose is to obtain the analytical expressions for the
magnetooptical functions of semiconductor crystals (reflectivity,
transmissivity, absorption, and bulk magnetosusceptibility),
including a high number of Rydberg excitons, taking into account
the effect of anisotropic dispersion and the coherence of the
electron and hole with the radiation field, as well to calculate
the positions of excitonic resonances in the situation when
degeneracies of exciton states with different orbital and spin
angular momentum are lifted by magnetic field. Presented approach,
owing to application of the full form of Hamiltonian for excitons
in an external magnetic field, allows one to get so-called
positive shifts of resonances (connected with a linear dependence
on the field strength and quadratic exciton diamagnetic shifts.
Due to the specific structure of Cu$_2$O crystal,  particulary to
a small radius of Wannier excitons, it is justified to assume
infinite confinement potentials at the crystal surface. All these
factors result in complex pattern of spectra which, especially for
higher order of excitons, becomes even more intricate for states
with higher principal number \emph{n}.

In very recent paper Schweiner \emph{et al}
\cite{Schweiner.Symmetry} has pointed out that an external
magnetic field influences the system, reducing its cubic
symmetry and leads to the complex splitting of excitonic lines in
absorption spectra. They have solved numerically Schr\"{o}dinger
equation and then calculated oscillator strengths  including the
complete valence band structure into their considerations. The most
results for Rydberg excitons was
     concentrated on the energy values of the excitonic states (for example, \cite{Schweiner.Magnetoexcitons}).  We, in turn,
extend such approach  including the polaritonic effects.
This will be done by means of the so-called center-of-mass
     quantization approach \cite{Tuffigo}-\cite{Strong}. Thanks to this
     approach one can include into account the influence of the internal structure of the
     electromagnetic wave propagating in the crystal. It is
     important because in the experiments on the Rydberg excitons in
     Cu$_2$O (\cite{Kazimierczuk}, \cite{Thewes}, \cite{assmann})
     the crystal size in the propagation direction exceeds largely
     the wavelength. Finally, we will examine the influence of the
     effective mass anisotropy on the magnetooptic properties.

     The paper is organized as follows. In
     Sec.~\ref{density.matrix} we recall the basic equations of
     the RDMA and formulate the equations for the case when the
     constant magnetic field is applied. In Sec.~\ref{iteration}
     we describe an iteration procedure, which will be applied to
     solve a system of coupled integro-differential equations and
     finally obtain the magneto-optic functions. The second iteration step,
     from which the magneto-optic functions, including the
     polariton effects, will be calculated, is given in
     Sec.~\ref{magnetooptic}. The formulas, derived in this
     Section, are than applied in Sec~\ref{results}. to calculate the magneto-optic
     functions for a Cu$_2$O crystal, considered in
     Ref.~\cite{assmann}. Finally, in Sec.~\ref{conclusions} we
     draw conclusions of the model studied in this paper. The derivations
     of useful matrix elements and calculations with a lot of  technical details
     are established in Appendices.

\section{Density matrix formulation}\label{density.matrix}
Having in mind the above mentioned experiments on Rydberg
excitons, we will compute the linear response of a semiconductor
slab to a plain electromagnetic wave, whose electric field vector
has a component of the form
\begin{equation}
E_i(z,t)=E_{\rm in}\exp({\rm i}k_0z-{\rm i}\omega t),\qquad
k_0=\frac{\omega}{c},
\end{equation}
attaining the boundary surface of the semiconductor located at the
plane $z=0$. The second boundary is located at the plane $z=L$. In
the case of the examined Cu$_2$O crystals the extension will be of
the order 30 $\mu\hbox{m}$.

The electromagnetic wave is then reflected, transmitted and
partially absorbed. The wave propagating in the medium has the
form of polaritons, defined as joint field-medium excitations.
Polaritons are mixed modes of the electromagnetic wave and
discrete excitations of the crystal (excitons). Below we assume
the separation of the relative electron-hole motion with well
defined quantum levels and the center-of-mass (COM) motion which
interacts with the radiation field and produces the mixed modes.

In the RDMA all this processes are described by a set of the
so-called constitutive equations for the coherent amplitudes
$Y^\nu(\textbf{r}_e,\textbf{r}_h)$ of the electron-hole pair of
coordinates $\textbf{r}_h$ (hole) and $\textbf{r}_e$ (electron).
In the case of Cu$_2$O $\nu$ means $P, F, H,\ldots$ excitons. The
equations have been described, for example, in Refs.
\cite{Zielinska.PRB}, \cite{Zielinska.PRB.2016.b} and have the
form (se also \cite{StB87}, \cite{RivistaGC})
 \begin{equation}\label{constitutiveeqn}
 \dot{Y}(\textbf{R},\textbf{r})+({\rm
 i}/\hbar)H_{eh}{Y}(\textbf{R},\textbf{r})+(1/\hbar){\mit\Gamma}{Y}(\textbf{R},\textbf{r})=({\rm
 i}/\hbar)\textbf{M}(\textbf{r})\textbf{E}(\textbf{R}),
 \end{equation}
where  ${\bf R}$ jest is the excitonic center-of-mass coordinate,
$\textbf{r}=\textbf{r}_e-\textbf{r}_h$ the relative coordinate,
$\textbf{M}(\textbf{r})$ the smeared-out transition dipole
density, ${\bf E}({\bf R})$ is the electric field vector of the
wave propagating in the crystal, and $\mit\Gamma$ stands for the
dissipation processes. The smeared-out transition dipole density
${\bf M}({\bf r})$ is related to the bilocality of the amplitude
$Y$ and describes the quantum coherence between the macroscopic
electromagnetic field and the interband transitions. The two-band
Hamiltonian $H_{eh}$ includes the electron- and hole kinetic
energy terms, the electron-hole interaction potential and the
confinement potentials. When constant  fields, magnetic
 and electric,  are applied, the Hamiltonian has
the form

\begin{eqnarray}\label{BFmagnetichamiltonian1}
&&\phantom{nucl}H=E_{g}+\frac{1}{2m_e} \left({\bf p}_e-e
\frac{{\bf r}_e \times {\bf B}}{2}\right)^2 + \frac{1}{2m_{hz}}
\left({\bf p}_h + e \frac{{\bf r}_h \times {\bf B}}{2}\right)_z^2
\nonumber\\ & &+ \frac{1}{2m_{h\parl}} \left( {\bf p}_h+e
\frac{{\bf r}_h \times {\bf B}}{2}\right)_\parl^2 +e{\bf
F}\cdot({\bf r}_e-{\bf r}_h) + V_{\rm conf}({\bf r}_e,{\bf r}_h) -
\frac{e^2}{4\pi\epsilon_0\epsilon_b \vert{\bf r}_e - {\bf
r}_h\vert} ,
\end{eqnarray}
 \noindent
 {\bf B} is the magnetic field
vector, ${\bf F}$ the electric field vector,  $V_{\rm conf}$ are
the surface potentials for electrons and holes, $m_{hz}, m_{h
\parl}$ are the components of the hole effective mass tensor, and the
electron mass is assumed to be isotropic. Separating
 the exciton center-of-mass and relative motion, and considering the case when ${\bf B}\parl
 z$,$\textbf{F}=0$,
we transform the  Hamiltonian (\ref{BFmagnetichamiltonian1}) into
the form

\begin{eqnarray}\label{SL3Hamiltonian}
H&=&H_0+\frac{P_z^2}{2M_z} + \frac{{\bf P}_\parl^2}{2M_\parl} +
\frac{1}{8}\mu_\parl
\omega_c^2 \rho^2 + \frac{e}{2\mu_\parl'}B {\mathcal L}_z   \nonumber\\
& & -\frac{e}{M_\parl} {\bf P}_\parl\cdot \left( {\bf
r}_\parl\times{\bf B} \right)+V_{\rm conf}({\bf R},{\bf r}),
\end{eqnarray}
where $\omega_c={eB}/{\mu_\parl}$ is the cyclotron frequency,
\noindent the reduced mass $\mu_{\parl}'$ is defined as
\begin{equation}\label{reducedmuprim}
 \frac{1}{\mu_{\parl}'}=\frac{1}{m_e}-\frac{1}{m_{h\parl}},
\end{equation}
\noindent and $H_0$ is the two-band Hamiltonian for the relative
electron-hole motion, as  used in the papers \cite{Zielinska.PRB},
\cite{Zielinska.PRB.2016.b}. The operator ${\mathcal L}_z$ is the
\emph{z}-component of the angular momentum operator. We must solve
the constitutive equations with the above Hamiltonian to obtain
the polarization and finally the polariton modes.

 The coherent amplitude $Y$ define the
excitonic counterpart of the polarization
\begin{equation}\label{polarization}
\textbf{P}_{\rm exc}(\textbf{R})=2 \int {\rm d}^3 r~
\textbf{M}^*(\textbf{r})Y(\textbf{R},\textbf{r}),
\end{equation}
which is than used in the Maxwell field equation
\begin{equation}\label{Maxwell}
c^2\nabla_R^2
\textbf{E}-\underline{\underline{\epsilon}}_{\,b}\ddot{\textbf{E}}(\textbf{R})=\frac{1}{\epsilon_0}\ddot{\textbf{P}}_{\rm
exc}(\textbf{R}),
\end{equation}
with the use  of the bulk dielectric tensor
$\underline{\underline{\epsilon}}_{\,b}$ and the vacuum dielectric
constant $\epsilon_0$. In the present paper we solve the equations
(\ref{constitutiveeqn})-(\ref{Maxwell}) with the aim to compute
the magnetoooptical functions (reflectivity, transmission, and
absorption) for the case of Cu$_2$O.

In semiconductors like, for example, GaAs, when only a few lowest
excitonic states are excited, it is possible to solve the
polariton dispersion relation and to determine the amplitudes of
the polariton waves. Analogous methods cannot be applied in the
case of Rydberg excitons, where, as, for example, in Cu$_2$O, even
25 excitonic states are observed. There is a question what
approach is appropriate for such case. One of the possibilities is
to use the so-called exciton center-of-mass (COM) quantization. In
this approach it is assumed, that no electron- or hole separately
are confined within the semiconductor crystal, but their
center-of-mass \cite{Tuffigo}-\cite{Strong}. This approach is
justified for small-radius Wannier excitons, as is the case of
Cu$_2$O (about 1 nm), and certainly not appropriate for
semiconductors with large-radius excitons, like GaAs (about 15
nm). In the COM approach mostly infinite confinement potentials at
the crystal surfaces $z=0,L$ are assumed, therefore the
eigenfunctions and eigenvalues of the COM motion have the form
\begin{eqnarray}\label{COMeigenfunctions}
w_N(Z)&=&\sqrt{\frac{2}{L^*}}\sin\left(\frac{N\pi}{L^*}Z\right),\nonumber\\
W_N&=&\frac{\hbar^2}{2M_z}\frac{N^2\pi^2}{L^{*2}}=N^2 S,\\
S&=&\frac{\mu_{\parallel}}{M_z}\left(\frac {\pi a^*}{L^*}\right)^2
R^*,\nonumber
\end{eqnarray}
where $M_z$ is the total excitonic mass in the $z-$ direction,
$a^*$ is the excitonic radius, $R^*$ the excitonic Rydberg, and
$L^*$ the effective crystal size in the $z-$ direction.

Having the confinement functions, we look for a solution
\begin{equation}\label{YY} Y(Z,\textbf{r})=\sum\limits_{Nn\ell m}c_{Nn\ell
m}R_{n\ell m}(r)Y_{\ell m}(\theta,\phi)w_N(Z).
\end{equation}
where $R_{n\ell m}$ are the radial functions of an anisotropic
Schr\"{o}dinger equation \cite{Zielinska.PRB.2016.b}
\begin{eqnarray}\label{radialfinala}
R_{n\ell m}(r)&=&\left(\frac{2\eta_{\ell
m}}{na^*}\right)^{3/2}\frac{1}{(2\ell
+1)!}\sqrt{\frac{(n+\ell)!}{2n(n-\ell-1)!}}\nonumber\\
 &&\times\left(\frac{2\eta_{\ell m}r}{na^*}\right)^\ell
e^{-\eta_{\ell m}r/na^*}M\left(-n+\ell+1,2\ell+2,\frac{2\eta_{\ell
m}r}{na^*}\right)\\
&=&\left(\frac{2\eta_{\ell
m}}{na^*}\right)^{3/2}\sqrt{\frac{(n-\ell-1)!}{2n(n+\ell)!}}\left(\frac{2\eta_{\ell
m}r}{na^*}\right)^\ell\;
L_{n-\ell-1}^{2\ell+1}\left(\frac{2\eta_{\ell
m}r}{na^*}\right) e^{-\eta_{\ell m}r/na^*},\nonumber\\
&&\eta_{\ell m}=\int {\rm d}\Omega\frac{\vert Y_{\ell
m}\vert^2}{\sin^2\theta +(\mu_\parl/\mu_z)\cos^2\theta},\nonumber
\end{eqnarray}
and $E_{n\ell m}$ the corresponding eigenvalues
\begin{equation}\label{Enellm}
E_{n\ell m}=-\frac{\eta_{\ell m}^2}{n^2}R^*,
\end{equation}
 $M(a,b,z)$ being the confluent hypergeometric function in the
notation of \cite{Abramowitz}, and $L_n^p(x)$ are the Laguerre
polynomials.

\section{Iteration procedure}\label{iteration}
The electric field $E(Z)$ of the wave propagating in the crystal,
acting as a source in Eq. (\ref{constitutiveeqn}), must satisfy
the Maxwell equation (\ref{Maxwell}) which, for the wave
propagating in the $Z$ direction and with the harmonic time
dependence fulfils the propagation equation
\begin{equation}\label{Maxwellzet}
\frac{{\rm d}^2E}{{\rm
d}Z^2}+k_b^2E(Z)=-\frac{\omega^2}{c^2\epsilon_0}P_{\rm
exc}(Z),\quad k_b=\sqrt{\epsilon_b}\frac{\omega}{c},
\end{equation}
and $P_{\rm exc}(Z)$ is the excitonic part of the crystal
polarization
\begin{eqnarray}\label{Polarization}
P_{\rm exc}(Z)&=&2\int M^*(\textbf{r})Y(Z,\textbf{r}){\rm d}^3r.
\end{eqnarray}
The function $Y(Z,\textbf{r})$, with regard to
(\ref{constitutiveeqn}),(\ref{YY}), and for the Faraday
configuration, satisfies the equation
\begin{eqnarray}\label{basic}
&&\sum_{Nn\ell m}\left(E_g-\hbar\omega +W_N+E_{n\ell
m}+m\frac{\mu_\parl}{\mu_\parl'}\gamma R^*-{\rm
i}{\mit\Gamma}+\frac{R^*}{4a^{*2}}\gamma^2r^2\sin^2\theta\right)c_{Nn\ell
m}R_{n\ell}(r)Y_{\ell
m}(\theta,\phi)w_N(Z)\nonumber\\
&&=\textbf{M}(\textbf{r})\textbf{E}(Z),
\end{eqnarray}
where $\gamma=\hbar\omega_c/2R^*$ is the dimensionless strength of
the magnetic field.  Even under applying the COM quantization, we
are left with
 a system of two coupled integro-differential equations
for the functions $Y$ and $E$. Having the field $E(Z)$ we can
determine the optical functions, reflectivity, transmissivity, and
absorption, by the relations
\begin{eqnarray}\label{RTdefinitions}
R&=&\left|\frac{E(0)}{E_{\rm in}}-1\right|^2,\quad
T=\left|\frac{E(L^*)}{E_{\rm in}}\right|^2,\nonumber\\
A&=&1-R-T,
\end{eqnarray}
where $E_{\rm in}$ is the amplitude of the normally incident wave.
The solution of the equations
(\ref{Maxwellzet})-(\ref{Polarization}), which yield the electric
field and the optical functions, can be obtained by several
methods. However, the methods, which were used for GaAs layers
\cite{CBT96}, or Quantum Dots (\cite{Schillak_epj},
\cite{PSCGC_2015} and references therein), cannot be applied for
the considered case of Rydberg excitons, as we discussed in our previous paper \cite{Zielinska.PRB},
\cite{Zielinska.PRB.2016.b}. Below we propose an iteration
procedure. The first step in this procedure is the solution of the
system of equations (\ref{basic}), where on the r.h.s. we put,
instead of the full solution $E(Z)$, its (known) homogeneous part
$E_{\rm hom}$, satisfying the equation
\begin{equation}\label{Maxwellzethom}
\frac{{\rm d}^2E}{{\rm d}Z^2}+k_b^2E(Z)=0,
\end{equation}
and the appropriate boundary conditions. It has the form
\begin{eqnarray}\label{Ehom}
E_{\rm hom}(Z)&=&E_{\rm in}\frac{2k_0f(L-Z)}{(k_b+k_0)W},\nonumber
\\
f(z)&=&e^{-{\rm i}k_bZ}+\frac{k_b-k_0}{k_b+k_0}e^{{\rm i}k_bZ},\\
W&=&e^{-{\rm i}k_bL}-\left(\frac{k_b-k_0}{k_b+k_0}\right)^2e^{{\rm
i}k_bL}.\nonumber
\end{eqnarray}
Inserting the above expression on the r.h.s. of the equations
(\ref{basic}), the following set of equations   will be obtained,  from
which the coefficients $c_{N n\ell m}$ can be determined

\begin{eqnarray}\label{basic1}
&&\sum_{Nn\ell m}\left(W_{Nn\ell
m}+\frac{R^*}{4a^{*2}}\gamma^2r^2\sin^2\theta\right)c_{Nn\ell
m}R_{n\ell m}(r)Y_{\ell
m}(\theta,\phi)w_N(Z)\nonumber\\
&&=\textbf{M}(\textbf{r})\textbf{E}_{\rm hom}(Z),\\
&&W_{Nn\ell m}=E_g-\hbar\omega +W_N+E_{n\ell
m}+m\frac{\mu_\parl}{\mu_\parl'}\gamma R^*-{\rm
i}{\mit\Gamma}.\nonumber
\end{eqnarray}
The expression for the dipole density, which should  be used in (\ref{basic}), has the
form: for $P$ excitons, $\ell=1$ \cite{Zielinska.PRB}
\begin{eqnarray}\label{gestoscwzbronione1}
{\bf M}^{(1)}({\bf r})&=&
\textbf{e}_r\,M_{10}\frac{r+r_0}{2r^2r_0^2}e^{-r/r_0}=\textbf{e}_r
M(r) =\textbf{i}M_{10}\frac{r+r_0}{4{\rm
i}r^2r_0^2}\sqrt{\frac{8\pi}{3}}\left(Y_{1,-1}-Y_{1,1}\right)e^{-r/r_0}\nonumber\\
&&+\textbf{j}M_{10}\frac{r+r_0}{4r^2r_0^2}\sqrt{\frac{8\pi}{3}}\left(Y_{1,-1}+Y_{1,1}\right)e^{-r/r_0}+\textbf{k}M_{10}\frac{r+r_0}{2r^2r_0^2}
\sqrt{\frac{4\pi}{3}}Y_{10}e^{-r/r_0},
\end{eqnarray}
and for $F$ excitons, $\ell=3$, in normalized (with respect to $r$)
form
\begin{eqnarray}\label{M3}
&&{\bf M}^{(3)}({\bf
r})=\textbf{i}M_{30}\frac{1}{r_0r^2}\left[\sqrt{\frac{3\pi}{7}}\left(Y_{31}-Y_{3-1}\right)
-\sqrt{\frac{5\pi}{7}}\left(Y_{33}-Y_{3-3}\right)\right]e^{-r/r_0}\nonumber\\
&&+\textbf{j}\frac{M_{30}}{r^2r_0}\left\{-{\rm
i}\left[\sqrt{\frac{3\pi}{7}}\left(Y_{31}+Y_{3-1}\right)-\sqrt{\frac{5\pi}{7}}\left(Y_{33}+Y_{3-3}\right)\right]\right\}e^{-r/r_0}\\
&&+\textbf{k}\frac{M_{30}}{r^2r_0}\left[12\sqrt{\frac{\pi}{7}}Y_{30}e^{-r/r_0}\right].\nonumber
\end{eqnarray}

 Having in mind the experiments by
A{\ss}mann et al \cite{assmann}, we consider the field \textbf{B}
as perpendicular to the crystal surface, ($\parl z$), and the wave
 propagating in the
$z$ direction and characterized by the electric field \textbf{E}.
The wave is assumed linearly polarized, $\textbf{E}=(E_x,0,0)$,
with the component $E_x=E_{\rm hom}(Z)$. Thus we take the $x$
components of the densities \textbf{M} defined in
(\ref{gestoscwzbronione1}) and (\ref{M3})
\begin{eqnarray}\label{Mx1}
M^{(1)}_x({\bf r})&=&M_{10}\frac{r+r_0}{4{\rm
i}r^2r_0^2}\sqrt{\frac{8\pi}{3}}\left(Y_{1-1}-Y_{11}\right)e^{-r/r_0},\\
\label{Mx3}M^{(3)}_x({\bf r})&=&
M_{30}\frac{1}{r_0r^2}\left[\sqrt{\frac{3\pi}{7}}\left(Y_{31}-Y_{3-1}\right)
-\sqrt{\frac{5\pi}{7}}\left(Y_{33}-Y_{3-3}\right)\right]e^{-r/r_0},
\end{eqnarray}
 with the coherence radius $r_0$ (see \cite{Zielinska.PRB} for the discussion about $r_0$). Using the above formulas, together with the expression
 (\ref{Ehom}) for  homogenous field $E_{\rm hom}$, we
 obtain, in the first step of iteration, a system of equations:
\begin{eqnarray}
&&\sum\limits_{n\ell m}\biggl[E_g-\hbar\omega +W_N+E_{n\ell
m}+m\frac{\mu_\parl}{\mu_\parl'}\gamma R^*-{\rm
i}{\mit\Gamma}\nonumber\\
&&+\frac{R^*}{4a^{*2}}\gamma^2r^2\sin^2\theta\biggr]c_{Nn\ell
m}R_{n\ell m}(r)Y_{\ell m}(\theta,\phi)w_N(Z)=M_xE(Z),
\end{eqnarray}
 By appropriate integration and
making use of the orthonormality of the eigenfunctions, the
equations for the expansion coefficients obtain the form
\begin{eqnarray}\label{coefficients_1}
&&W_{N_1n_1\ell_1m_1}c_{N_1n_1m_1\ell_1}+
\frac{R^*}{4a^{*2}}\gamma^2\sum\limits_{n\ell m}\langle
R_{n_1\ell_1 m_1}\vert r^2\vert R_{n\ell m}\rangle\langle
Y_{\ell_1m_1}\vert\sin^2\theta\vert Y_{\ell m}\rangle c_{N_1n\ell
m}\nonumber\\
&&=\langle  w_{N_1}\vert E_x(Z)\rangle\langle R_{n_1\ell_1
m_1}Y_{\ell_1m_1}\vert M_x\rangle.
\end{eqnarray}
 Introducing the notation
\begin{eqnarray}\label{definitionVn}
V^{(nn_1)}_{\ell\ell_1m m_1}&=&\frac{R^*}{4a^{*2}}\gamma^2\langle
R_{n_1\ell_1m_1}\vert r^2\vert R_{n\ell m}\rangle\langle
Y_{\ell_1m_1}\vert\sin^2\theta\vert Y_{\ell m}\rangle,
\end{eqnarray}
we put the equations (\ref{coefficients_1}) into the form
\begin{eqnarray}
&&W_{N_1n_1\ell_1m_1}c_{n_1m_1\ell_1}+ \sum\limits_{n\ell
m}V^{(nn_1)}_{\ell\ell_1m m_1} c_{N_1n\ell m} =\langle
w_{N_1}\vert E_x\rangle\langle R_{n_1\ell_1 m_1}Y_{\ell_1m_1}\vert
M_x\rangle.
\end{eqnarray}
As in the case of previously discussed electro-optic properties
\cite{Zielinska.PRB.2016.b}, we take into account only the states
$n=n_1$. This assumption is justified by the fact, that the
diamagnetic shift, related to these matrix elements, is much
smaller than the Zeeman splitting. Now we obtain the equations
\begin{eqnarray}\label{basic_equations}
&&W_{Nn\ell m}c_{Nn\ell m}+ V^{(n)}_{\ell\ell m} c_{Nn\ell m}
+V^{(n)}_{\ell\ell+2m} c_{Nn\ell+2 m}+V^{(n)}_{\ell\ell-2m}
c_{Nn\ell-2 m}\nonumber\\
&&\phantom{aaaa}=\langle  w_{N}\vert E_x\rangle\langle R_{n\ell_1m_1}Y_{\ell m}\vert M_x\rangle,\\
&&V^{(n)}_{\ell\ell
m}=\frac{R^*}{4}\gamma^2\frac{(\ell^2+\ell+m^2-1)}{(2\ell-1)(2\ell+3)}\left(\frac{n}{\eta_{\ell
m}}\right)^2[5n^2+1-3\ell(\ell+1)],\nonumber\\
&&V^{(n)}_{\ell\ell+2m}=-\frac{R^*}{4}\gamma^2\sqrt{\frac{(\ell+2-m)(\ell+1-m)(\ell+m+1)(\ell+m+2)}{(2\ell+1)
(2\ell+3)^2(2\ell+5)}}\int\limits_0^\infty{\rm d}\rho
\rho^4\,R_{n\ell}R_{n\ell+2},\nonumber\\
&&V^{(n)}_{\ell\ell-2m}=-\frac{R^*}{4}\gamma^2\sqrt{\frac{(\ell-m)(\ell-m-1)(\ell+m)(\ell+m-1)}{(2\ell-3)
(2\ell-1)^2(2\ell+1)}}\int\limits_0^\infty{\rm d}\rho
\rho^4\,R_{n\ell m}R_{n\ell-2m}.\nonumber
\end{eqnarray}
 The effect of overlapping of
different states is included in the resonant denominators. The
detailed form for the coefficients $c_{Nn\ell m}, \langle
R_{n_1\ell_1 m_1}Y_{\ell_1m_1}\vert M_x\rangle, \langle E_x\vert
w_{N}\rangle$, and the derivation of the matrix elements
$V^{(n)}_{\ell\ell m}$ is given in Appendices A and B. The equations
(\ref{basic_equations}) are the basic equations in the presented
paper, which will be used in the numerical calculations of the
optical functions.
\begin{figure}[h]
\includegraphics[width=0.6\linewidth]{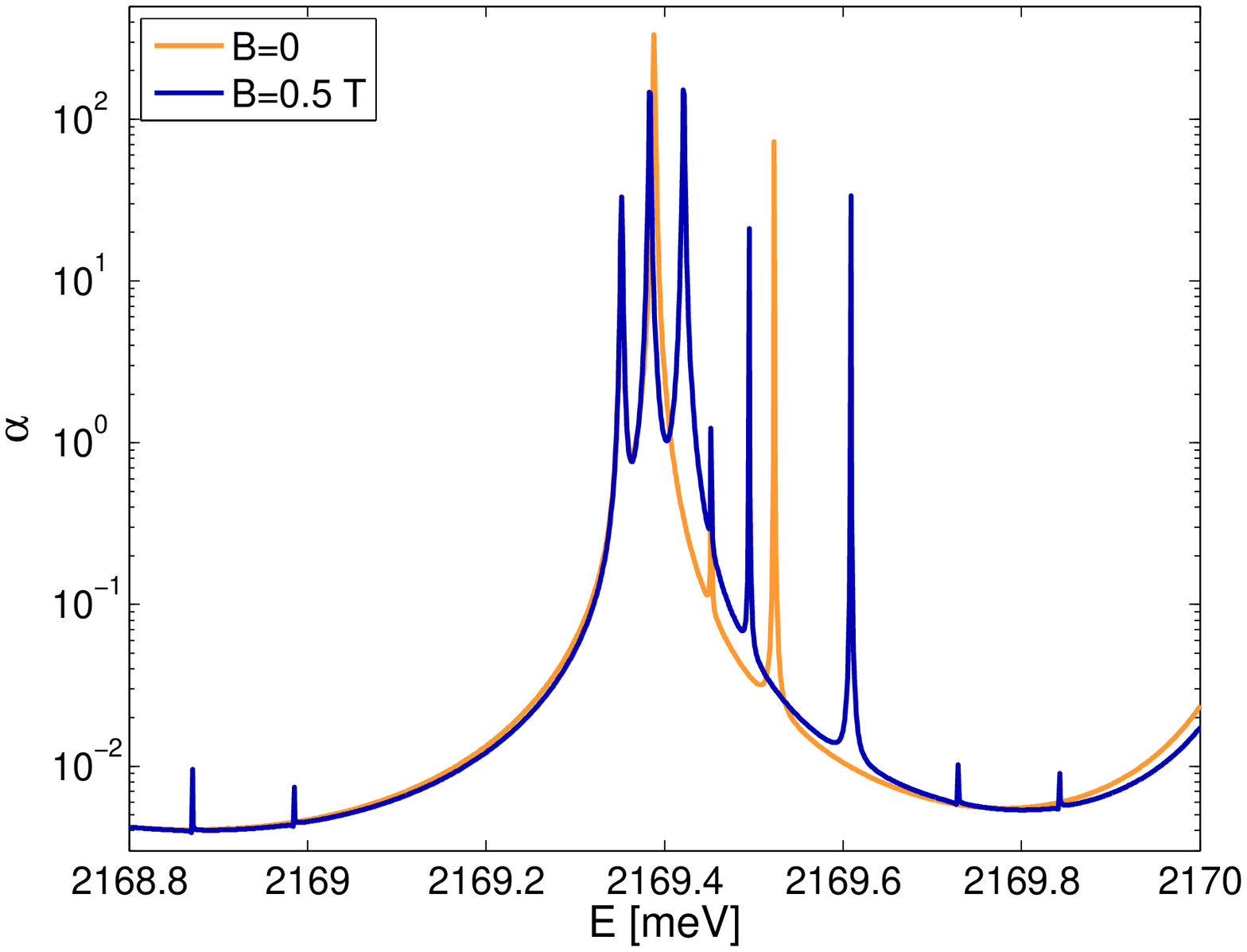}
\caption{Changes in absorption spectrum due to the applied
magnetic field for n=4 exciton.}\label{B0B1}
\end{figure}

\begin{figure}[h]
\includegraphics[width=0.6\linewidth]{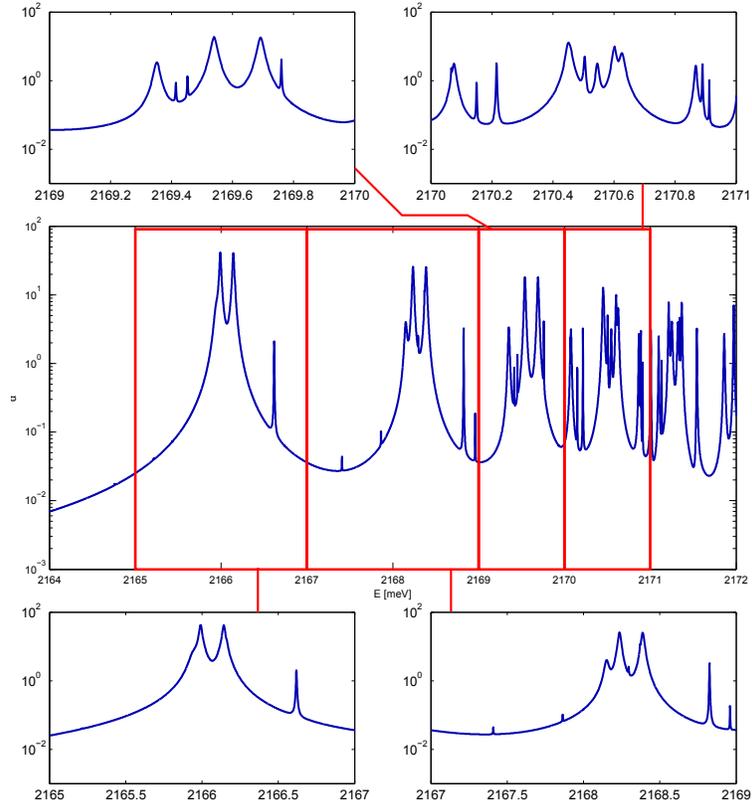}
\caption{The bulk magnetoabsorption of Cu$_2$O crystal calculated
from the imaginary part of the bulk susceptibility
(\ref{bulksusceptibility23}) for $B$=2T. Insets show the detailed spectra around the excitonic lines corresponding to n=4, 5, 6, 7.} \label{a_2T}
\end{figure}
\begin{figure}[h]
\includegraphics[width=0.65\linewidth]{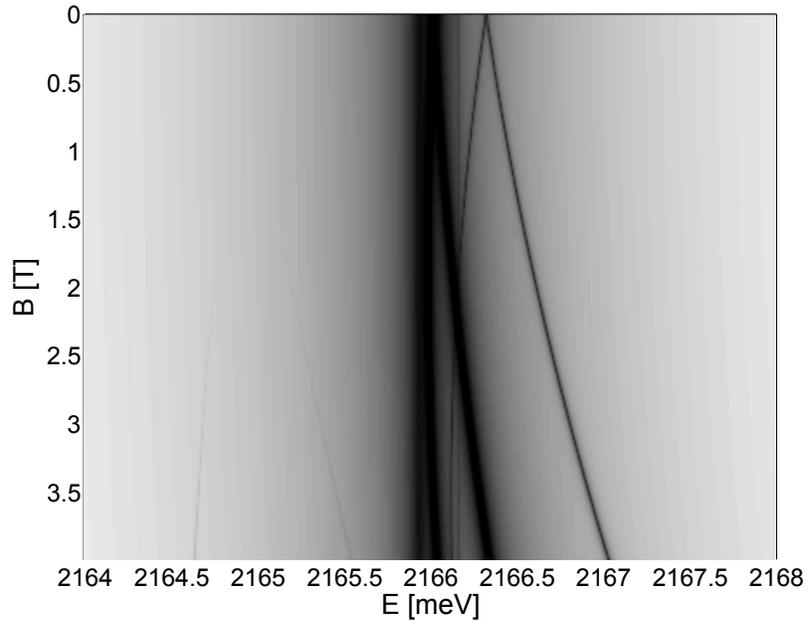}
\caption{Absorption spectrum in the energetic region of $n=4$
excitonic state as a function of the applied magnetic field
strength.}\label{polen4}
\end{figure}
\begin{figure}[h]
\includegraphics[width=0.65\linewidth]{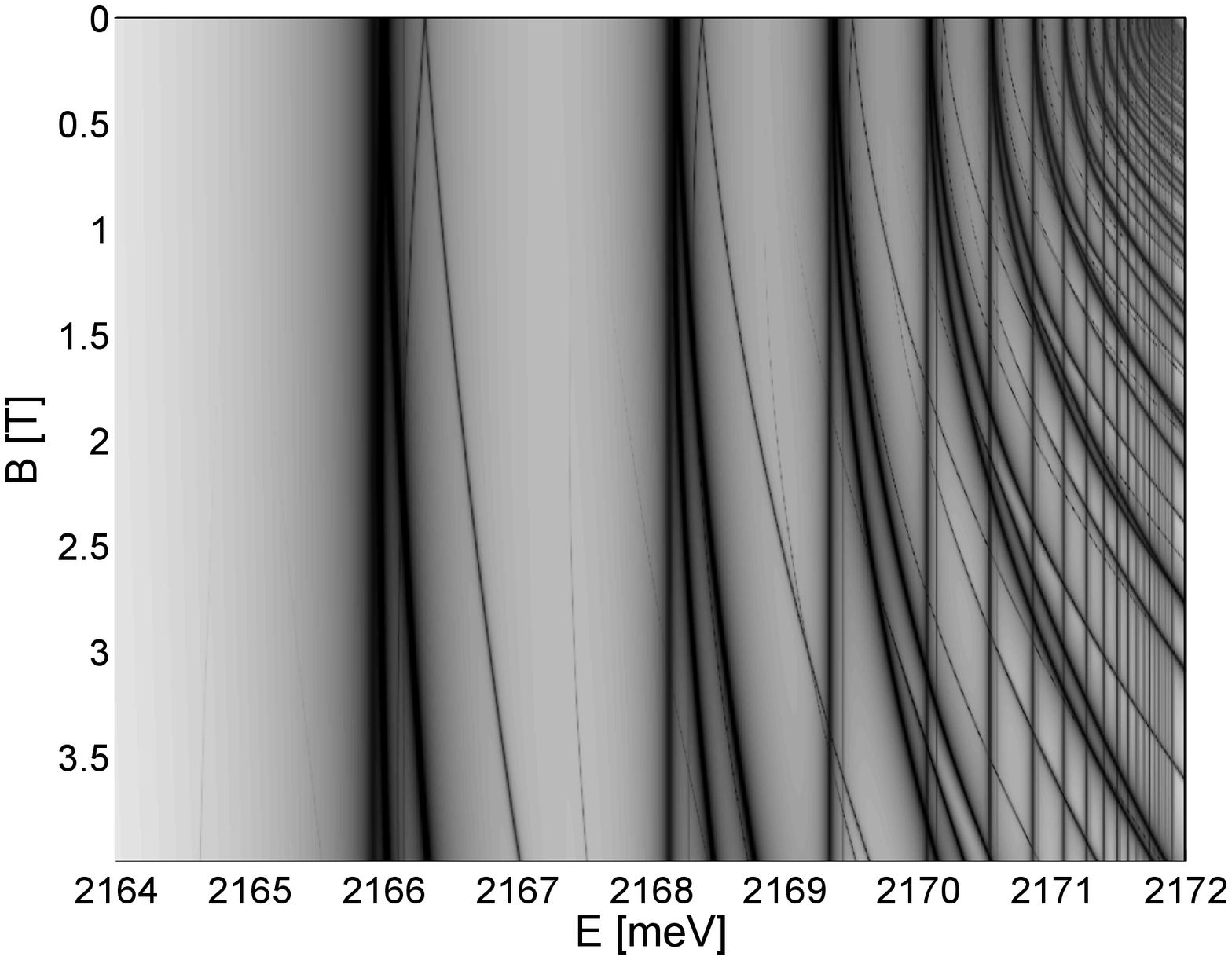}
\caption{Te same as in Fig~\ref{polen4}, in the energetic region
of $n=4-25$ excitonic states.}\label{n25}
\end{figure}
\section{Second iteration step - magneto-optic
functions}\label{magnetooptic}

\begin{figure}[h]
\includegraphics[width=0.65\linewidth]{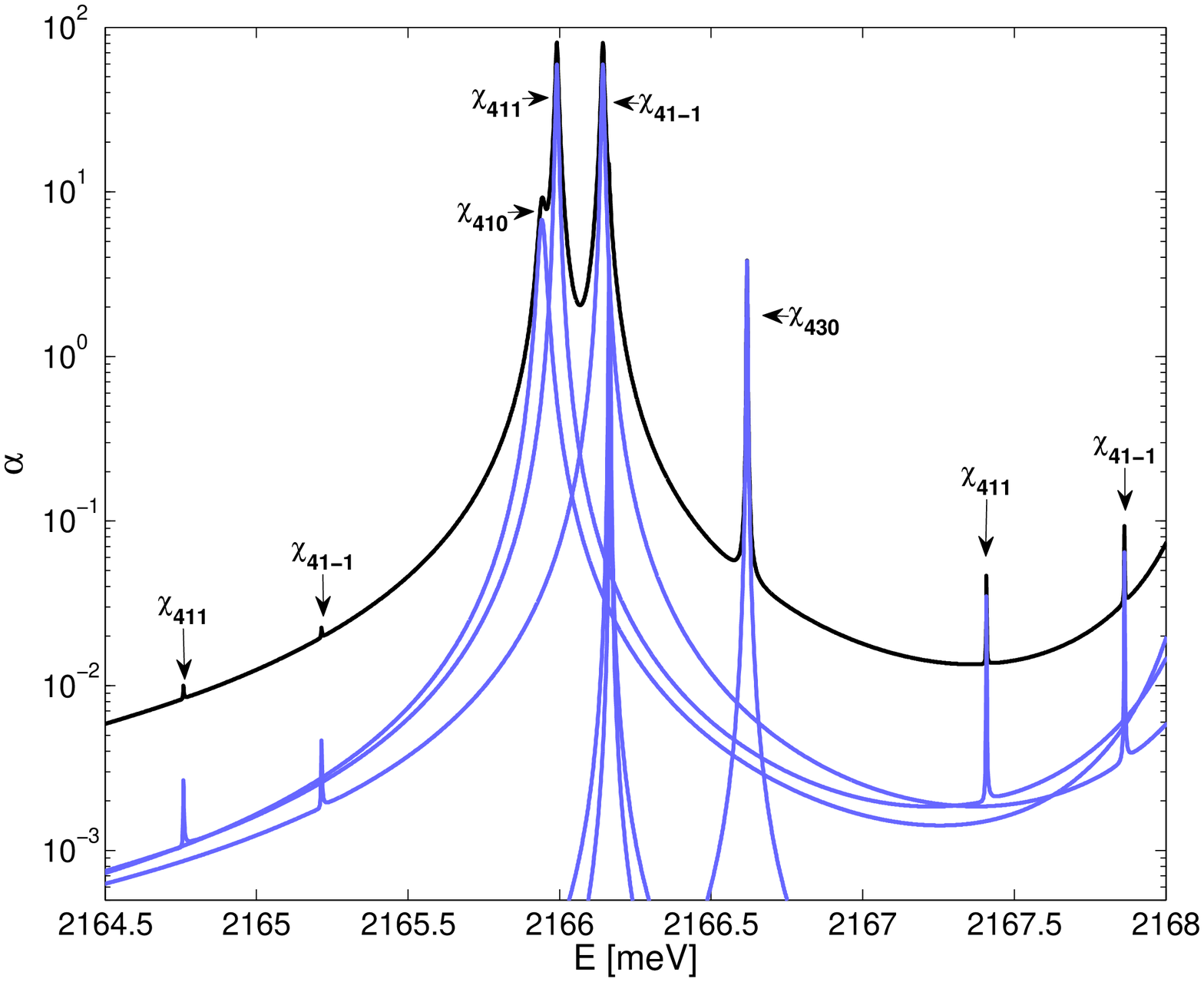}
\caption{The bulk magnetoabsorption of Cu$_2$O crystal, individual
excitonic resonances are identified.}\label{identification}
\end{figure}

\begin{figure}[h]
\includegraphics[width=0.65\linewidth]{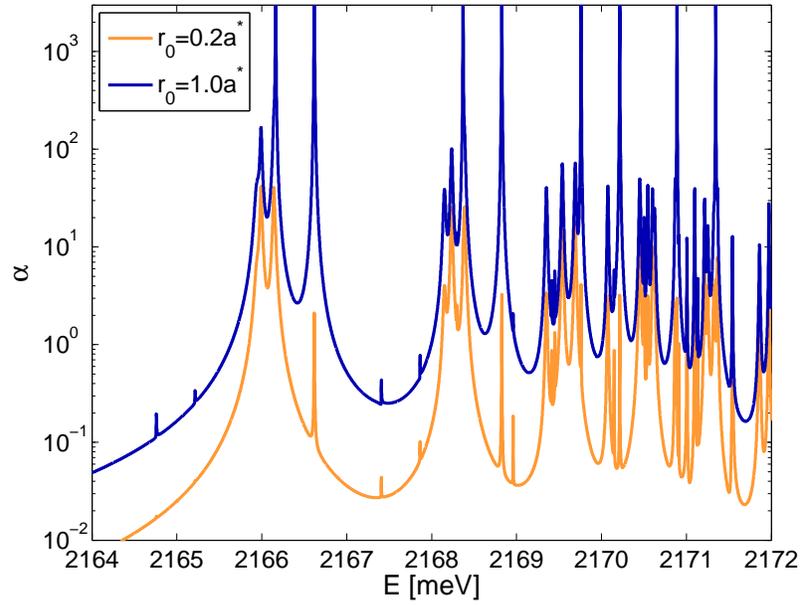}
\caption{The influence of the choice of the coherence radius $r_0$
on the magnetoabsorption spectra of Cu$_2$O crystal.}\label{R0}
\end{figure}
\begin{figure}[h]
\includegraphics[width=0.65\linewidth]{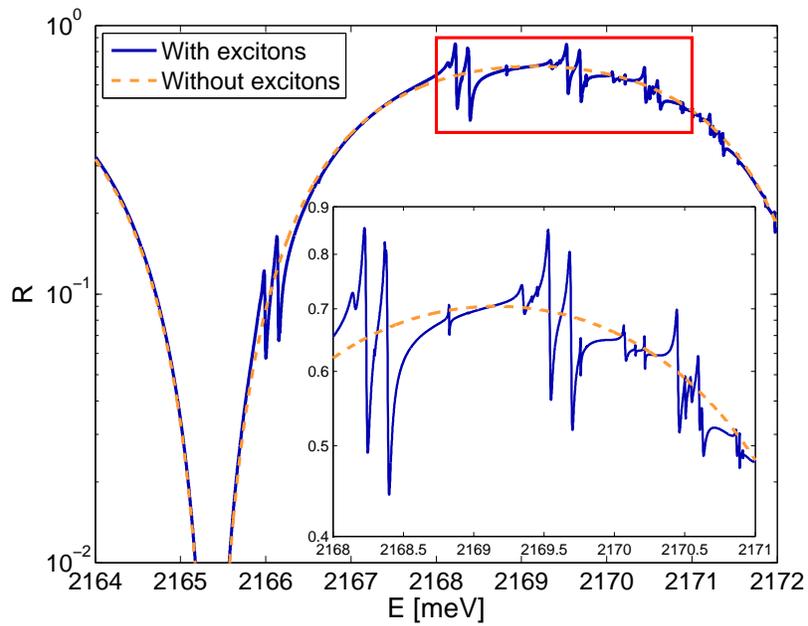}
\caption{The magneto-reflection coefficient of Cu$_2$O crystal of
thickness 30 $\mu\hbox{m}$, when the magnetic field 2T is applied.
The dashed curve (labelled without excitons) corresponds to
Fabry-Perot reflection. Inset - reflection spectrum for n=5, 6, 7 exciton.}\label{R302T}
\end{figure}

\begin{figure}[h]
\includegraphics[width=0.65\linewidth]{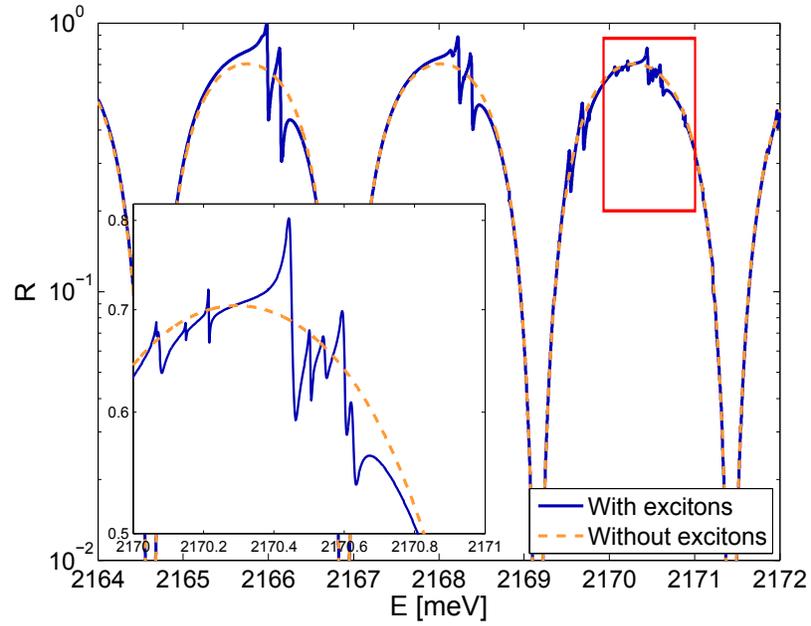}
\caption{The same as in Fig.~\ref{R302T}, for crystal thickness
100 $\mu\hbox{m}$. Inset - reflection spectrum for n=7 exciton.}\label{r100}
\end{figure}
\begin{figure}[h]
\includegraphics[width=0.65\linewidth]{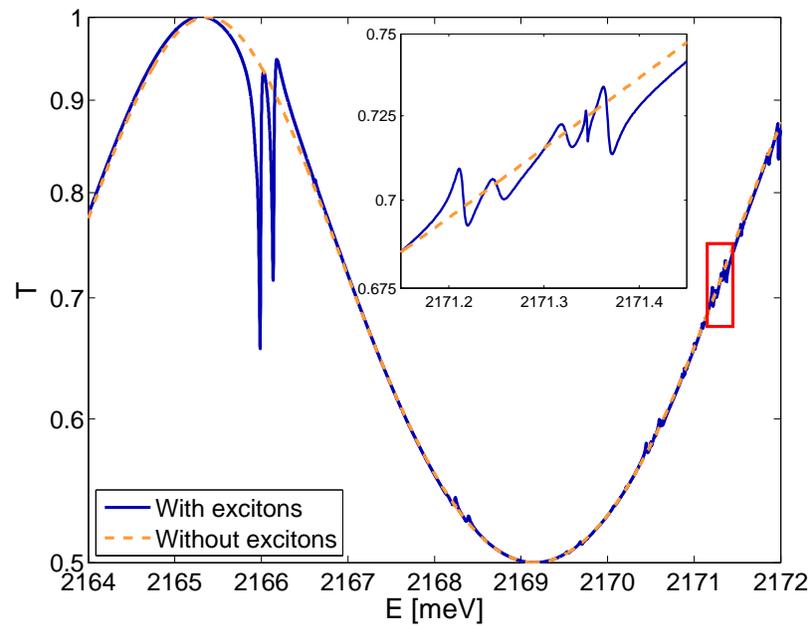}
\caption{The transmissivity coefficient of Cu$_2$O crystal of
thickness 30 $\mu\hbox{m}$, when the magnetic field 2T is applied.
The dashed curve (labelled without excitons) corresponds to
Fabry-Perot transmissivity. Inset - transmission spectrum for n=7 exciton.}\label{T302T}
\end{figure}
\begin{figure}[h]
\includegraphics[width=0.65\linewidth]{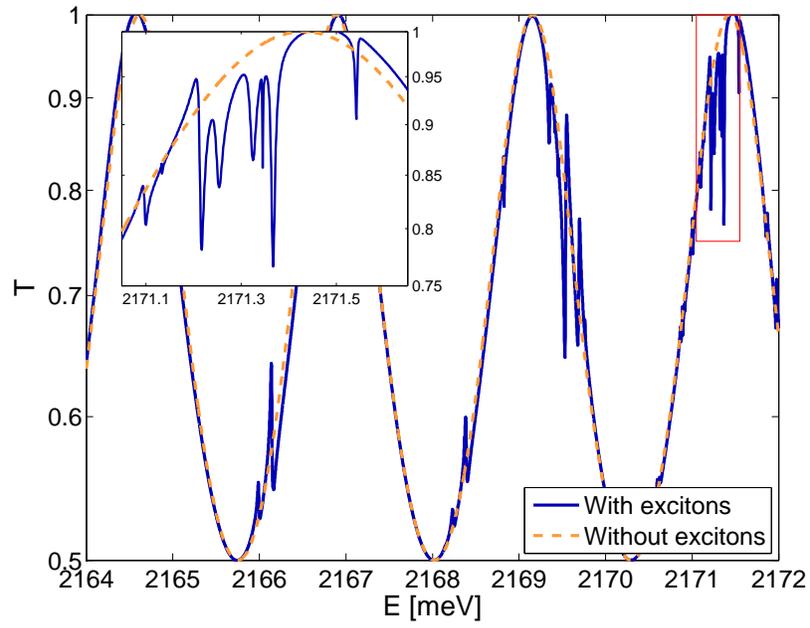}
\caption{The same as in Fig.~\ref{T302T}, for crystal thickness
100 $\mu\hbox{m}$. Inset - transmission spectrum for n=7 exciton.}\label{T1002T}
\end{figure}

In the first iteration step we have
  computed the coefficients
$c_{Nn\ell m}$. Having them, we determine the amplitude
$Y(Z,\textbf{r})$ from Eq. (\ref{YY})  and the excitonic
polarization from the Eq. (\ref{Polarization}), obtaining
\begin{eqnarray}\label{Polaryzacja_1}
P_{\rm exc}(Z)&=&2\int M^*(\textbf{r})Y(Z,\textbf{r}){\rm d}^3r
=2\int M^*(\textbf{r})\sum\limits_{Nn\ell m}c_{Nn\ell m}R_{n\ell
m}Y_{\ell m}w_N(Z){\rm d}^3r
=\sum\limits_NP_N w_N(Z),\nonumber\\
P_N&=&2\sum\limits_{n\ell m}c_{Nn\ell m}\langle M^*\vert R_{n\ell
m}Y_{\ell m}\rangle=\epsilon_0\epsilon_b\Delta_{LT}^{(P)}\langle
w_N\vert E_{\rm hom}(Z)\rangle\sum_{n\ell m}\chi_{N n\ell m}\\
&&=\epsilon_0\epsilon_b\Delta_{LT}^{(P)}I_NE_{\rm
hom}(0)\sum_{n\ell m}\chi_{N n\ell m},\nonumber
\end{eqnarray}
\noindent where $I_N$ are defined in Eq. (\ref{ien}), and $\chi_{N
n\ell m}$ in Eqn. (\ref{chin23}, \ref{chingeq4}). The so obtained
polarization will be used as a source in the Maxwell equation
(\ref{Maxwellzet}). As is known, the solution of a nonhomogeneous
differential equation is composed of two parts, which are the
solution of a homogeneous equation and of the nonhomogeneous one:
\begin{equation}\label{poleejednplus}
E(Z)=E_{\rm hom}(Z)+E_{\rm nhom}(Z),
\end{equation}
\noindent where the homogeneous part satisfies the equation
(\ref{Maxwellzethom}). The nonhomogeneous part will be obtained by
means of the appropriate Green function, satisfying the equation

\begin{equation}
\frac{{\rm d}^2}{{\rm
d}Z^2}G^E(Z,Z')+k_b^2G^E(Z,Z')=-\delta(Z-Z'),
\end{equation}
\noindent and having the form (for example,~\cite{RivistaGC})
\begin{eqnarray}\label{Greenelectric}
G^E(Z,Z')&=&\frac{\rm i}{2k_bW}\left(e^{-{\rm i}k_bZ^<}+\frac{k_b-k_0}{k_b+k_0}e^{{\rm i}k_bZ^<}\right)\nonumber\\
&\times&\left(\frac{k_b-k_0}{k_b+k_0}e^{{\rm i}k_bL-{\rm
i}k_bZ^>}+e^{-{\rm i}k_bL+{\rm i}k_bZ^>}\right),
\end{eqnarray}
\noindent where $Z^<=\hbox{min} (Z,Z'), Z^>=\hbox{max}(Z,Z')$.
Using the above Green's function we obtain the nonhomogeneous part
in the form
\begin{equation}\label{Enonhomogeneous}
E_{\rm
nhom}(Z)=\frac{k_0^2}{\epsilon_0}\int\limits_0^LG^E(Z,Z')P_{\rm
exc}(Z'){\rm d}Z'.
\end{equation}
The equations (\ref{poleejednplus}), (\ref{Greenelectric}), and
(\ref{Enonhomogeneous}) give the total electric field of the wave
propagating in the crystal, from which the reflectivity $R$ and
transmissivity $T$ are obtained. They have the form
\begin{eqnarray}\label{reflectionR}
R&=&\left|\frac{E(0)}{E_{\rm in}}-1\right|^2\\
&&=R_{\rm FP}\Biggl|1+{\rm
i}\left(\frac{k_bL}{2}\right)\frac{(1-r_\infty^2)(1-r_\infty
e^{{\rm i}\Theta})^2(1-r_\infty^2e^{{\rm
i}\Theta})}{r_\infty(1-e^{{\rm
i}\Theta})}\sum\limits_N\left(\frac{I_N^2}{L}\right)\sum\limits_{n\ell
m}\Delta_{LT}^{(P)}\chi_{N n\ell m}\Biggr|^2,\nonumber\\
\label{transmission} T&=&\left|\frac{E(L)}{E_{\rm
in}}\right|^2=T_{\rm FP}\left|1+{\rm
i}\left(\frac{E}{2E_L}\right)\frac{(1-r_\infty e^{{\rm
i}\Theta})}{(1-r_\infty^2e^{{\rm
i}\Theta})}\sum\limits_N\left(\frac{I_N^2}{L}\right)\sum\limits_{n\ell
m}\Delta_{LT}^{(P)}\chi_{N n\ell m}\right|^2,
\end{eqnarray}
where $\Theta=2k_bL$, $r_\infty$ is defined by
\begin{equation}
r_\infty=\frac{k_0-k_b}{k_0+k_b},
\end{equation}
$I_n^2/L$ are given in Eq. (\ref{INsquare}), and $R_{\rm FP},
T_{\rm FP}$ are the well-known formulas for the Fabry-Perot normal
incidence reflectivity and transmissivity of a lossless dielectric
slab of thickness $L$ (see, e.g., \cite{Klingshirn}),
\begin{eqnarray}\label{FabryPerot}
R_{\rm FP}&=&\frac{F\sin^2(\Theta/2)}{1+F\sin^2(\Theta/2)},\quad
T_{\rm FP}=\frac{1}{1+F\sin^2(\Theta/2)},\nonumber\\
F&=&\frac{4r_\infty^2}{(1-r_\infty^2)^2}.
\end{eqnarray}
The derivation of the formulas (\ref{reflectionR}) and
(\ref{transmission}) is given in Appendix E.

\section{Results}\label{results}
We have performed numerical calculations of magneto-optical
functions (absorption, reflectivity, and transmissivity) for the
Cu$_2$O crystal having in mind the experiments by A{\ss}mann
\emph{et al}~\cite{assmann}. Considering the specific properties
of Cu$_2$O and the crystal dimension large compared to the exciton
Bohr radius, we observe that the COM quantized energies $W_N$
(\ref{COMeigenfunctions}) are small compared to the remaining
components of the excitonic energies $W_{Nn\ell m}$
(\ref{basic1}). Therefore, in the first approximation, we can
neglect them in the formulas defining the coefficients $c_{Nn\ell
m}$ (Appendix~\ref{Appendix A}) and in expressions $\chi_{Nn\ell
m}$ (\ref{chin23},\ref{chingeq4}), and calculate the bulk
magneto-susceptibility
\begin{eqnarray}\label{bulksusceptibility23}
\chi(\omega)&=&\epsilon_b\Delta_{LT}^{(P)}\left[\chi_{N211}+\chi_{N21-1}+\chi_{N311}+\chi_{N31-1}+\chi_{Nn3\pm
3}+\tilde{\chi}_{Nn1\pm 1}+\tilde{\chi}_{Nn3\pm 1}\ldots\right],
\end{eqnarray}
Using the formula
$\alpha=(\omega/c)\hbox{Im}\,\sqrt{\epsilon_b+\chi}$ we have
calculated the magneto-absorption, taking into account the lowest
$n=2-10$ excitonic states. The parameters we used are the energies
$E_{n\ell m}$, the gap energy $E_g$, the L-T energy
$\Delta_{LT}^{(P)}$, and the dissipation parameter $\mit\Gamma$.
We have used the electron and hole effective masses: $m_e=1.01,
m_{h\parallel}=0.5587, \mu_\parallel=0.3597,
\mu_z=0.672\;\hbox{(the masses in free electron mass)}\; m_0$,
from them calculated the reduced mass $\mu_{\parallel}'=-1.25$,
which gives ${\mu_{\parallel}}/{\mu_{\parallel}'}\approx -0.2876.$
Since the LT splitting for \emph{P} excitons is not known, we have
established a relation between the known $\Delta_{LT}$ for
\emph{S} excitons and the quantity $\Delta_{LT}^{(P)}$ for
\emph{P} excitons. First, using the bulk dispersion
\begin{equation}
\frac{c^2k^2}{\omega^2}=\epsilon_b+\frac{2}{\epsilon_0}\int {\rm
d}^3r M^* Y,
\end{equation}
for $k=0, n=2$, we establish the relation between the splitting
and the dipole matrix element
\begin{eqnarray}
\epsilon_b\Delta_{LT}^{(P)}&=&\left.\frac{2}{\epsilon_0}\left|I_1+I_2\right|^2\left[\frac{1}{W_{111}}+\frac{1}{W_{11-1}}\right]\right|_{B=0},\nonumber\\
\left|M_{10}\right|^2&=&\frac{4\epsilon_0\epsilon_ba^{*3}\Delta_{LT}^{(P)}}{\pi
(r_0/a^*)^2\eta_{11}^5},
\end{eqnarray}
where $I_1,I_2$ are defined in Eq.(\ref{I_1},\ref{I_2}). Using an
analogous expression for   for \emph{S} excitons
(\cite{Zeitschrift})
$$\left|M_{10}\right|^2=\frac{\pi\epsilon_0\epsilon_ba^{*3}\Delta_{LT}^{(S)}}{\eta_{00}^3}$$
one can determine $\Delta_{LT}^{(P)}$ as function of
$\Delta_{LT}^{(S)}$ \begin{eqnarray}\label{deltap}
\Delta_{LT}^{(P)}&=&\frac{\pi^2}{4}\left(\frac{r_0}{a^*}\right)^2\frac{\eta_{11}^5}{\eta_{00}^3}\Delta_{LT}^{(S)}.
\end{eqnarray}
The energies $W_{Nn\ell m}$ were obtained from the relations
(\ref{Enellm}), and (\ref{basic1}) (without $W_N$) with the
effective Rydberg energy $R^*$. We have used the values
$E_g=2172~\hbox{meV}, R^*=86.981~\hbox{meV}$,
$\Delta^{(S)}_{LT}=10~\mu\hbox{eV}$ which is common value in
available literature, $\mu_{\parallel}/\mu_z=0.5351$, and
phenomenological value of damping ${\mit\Gamma}=0.1~\hbox{meV}$.
We have restricted the upper limit of energy for our numerical
illustrations to $2172~\hbox{meV}$, which determines the band gap;
above this energy one could take into account effects which follow
from interaction with the continuous spectrum. The results for the
absorption, which seem the most important, are reported in
Figs.~\ref{B0B1}-\ref{R0}. In Fig.~\ref{B0B1} we observe how the
applied magnetic field changes the spectra. The positions of
absorption maxima are changed due to the diamagnetic shift, and
the Zeeman splitting occurs. The same effects, for a larger number
of excitonic states and stronger field, are displayed in Fig.~\ref{a_2T}. For the
principal quantum number $n\geq 4$ the effects due to both
\emph{P} and \emph{F} excitons, for example, the overlapping of
states, are observed. This is shown in Fig.~\ref{polen4} for the
$n=4$ exciton state, where the dependence of the absorption on the
applied field strength is shown. The Zeeman splitting is clearly visible and the lines are shifted towards higher energy with increasing field strength. Moreover, some absorption lines become visible only for sufficiently strong magnetic field. Our calculation scheme can be extended to higher principal quantum number; the absorption spectrum for n=4-25 excitonic states is shown on the Fig. \ref{n25}. In Fig.~\ref{identification} we show that our method allows to
identify the excitonic states. In the RDMA approach we consider
the role of the coherence of the radiation field with excitons,
which enters \emph{via} the smeared dipole density
\textbf{M}(\textbf{r}) and its parameters $\textbf{M}_0, r_0$. The
influence of the coherence radius $r_0$ is illustrated in
Fig.~\ref{R0}. One can observe that an increase of $r_0$ causes an
increase of absorption, which is due to the related increase of
the L-T splitting and oscillator strength, see Eq. (\ref{deltap}).
The COM quantization approximation, used in our calculations,
allowed to calculate the reflection and transmission spectra,
taking into account both the microscopic electronic excitations
(excitons) and their interaction with the radiation field,
resulting in creation of polariton modes. We have obtained
analytical expressions for the reflection coefficient $R$ and the
transmissivity $T$, and their shapes are shown in
Figs.~\ref{R302T}-\ref{T1002T}. The calculation has been performed up to N=350 and for excitonic states with principal number n=4-7. One can observe the overlapping of
the Fabry-Perot modes, typical for a dielectric slab, with
excitonic resonances. Notably, the Fabry-Perot interference has the dominating effect, strongly affecting the reflection and transmission coefficient over the whole energy range. Conversely, the variation of reflection coefficient due to the excitonic resonances has a comparably small amplitude, but is quickly varying with energy. In the Fig. \ref{R302T} inset, one can see that these two effects are readily separable.

\section{Conclusions}\label{conclusions}
The main results of our paper can be summarized as follows. We
have proposed a procedure based on the RDMA approach that allows
to obtain analytical expressions for the magneto-optical functions
of semiconductor crystals including high number Rydberg excitons.
Our results have general character because arbitrary exciton
angular momentum number and arbitrary applied field strength are
included. We have chosen the example of cuprous oxide, inspired
by the recent experiment by A{\ss}mann \emph{et
al}~\cite{assmann}. We have calculated the magneto-optical
functions (susceptibility, absorption, reflection, and
transmission),
  obtaining a {fairly} good agreement between the calculated and the experimentally observed
  spectra.

 {As each method using  iterative procedure presented approach is a kind of approximation.
Although we have solved the problem of excitons in semiconductor
when an external magnetic field is applied, but we do not include
the interaction between states with different \emph{n} into the
Hamiltonian we have used. It should emphasized that presented
approach is analytical until the last step in which the
numerical code is used.}
  {The choice of dipole density model and therefore the oscillator strengths,
which is  intricate function of  free parameters has an impact on
accuracy of our calculations {and might be the source of
discrepancy between the experimental results}}.
  Our results  confirm the fundamental  peculiarity
of magneto-optical effects: shifting, splitting and, as a result
for higher excitonic states, mixing of spectral lines.
   In particular, we obtained the splitting of \emph{P} and \emph{F} excitons, with increasing number of peaks corresponding to the increasing state number.
     All these interesting
features of excitons with high $n$ number which are examined
     and discussed  on the basis on our theory might possibly provide deep insight into the nature of Rydberg excitons
     in solids and provoke their application to design all-optical flexible switchers and future implementation in
      quantum information processing.
\appendix
\section{Expansion coefficients $c_{Nn\ell m}$}\label{Appendix A}
Using the eigenfunctions (\ref{radialfinala}) and  transition
dipole densities (\ref{Mx1}), (\ref{Mx3}) we calculate the
expansion coefficients $c_{Nn\ell m}$. For $n=2,3$ we only hace
the coefficients related to the $P$ excitons:
\begin{eqnarray}
W_{N_1n1\pm 1}c_{N_1n1\pm 1}+V^{(n)}_{11\pm 1}c_{N_1n1\pm
1}&=&\langle  w_{N_1}\vert E_x\rangle\langle R_{211}Y_{1\pm
1}\vert
M_x^{(1)}\rangle,\nonumber\\
c_{N_1n1\pm 1}&=&\frac{\langle w_{N_1}\vert E_x\rangle\langle
R_{n11}Y_{1\pm 1}\vert M_x^{(1)}\rangle}{W_{N_1n1\pm
1}+V^{(n)}_{11\pm 1}}.
\end{eqnarray}
For $n\geq 4$ we must consider contributions of both excitons $P$
and $F$. For $n=4$ we have three equations
\begin{eqnarray}\label{cn4}
&&W_{N_141\pm 1}c_{N_141\pm 1}+V^{(4)}_{11\pm 1}c_{N_141\pm
1}+V^{(4)}_{13\pm 1}c_{N_143\pm 1}=\langle  w_{N_1}\vert
E_x\rangle
\langle R_{411}Y_{1\pm 1}\vert M^{(1)}_x\rangle,\nonumber\\
&&W_{N_143\pm 1}c_{N_143\pm 1}+V^{(4)}_{33\pm 1}c_{N_143\pm
1}+V^{(4)}_{31\pm 1}c_{N_141\pm 1}=\langle  w_{N_1}\vert
E_x\rangle
\langle R_{431}Y_{3\pm 1}\vert M^{(3)}_x\rangle,\\
&&W_{N_143\pm 3}c_{N_143\pm 3}+V^{(4)}_{33\pm 3}c_{N_143\pm
3}=\langle  w_{N_1}\vert E_x\rangle \langle R_{433}Y_{3\pm 3}\vert
M^{(3)}_x\rangle\nonumber.
\end{eqnarray}
The first two equations  can be put into the form
\begin{eqnarray}
&&a_{11}c_{N_141\pm 1}+a_{12}c_{N_143\pm 1}=b_1,\nonumber\\
&&a_{21}c_{N_141\pm 1}+a_{22}c_{N_143\pm 1}=b_2,\nonumber\\
&&a_{11}=W_{N_141\pm 1}+V^{(4)}_{11\pm 1},\\
&&a_{12}=V^{(4)}_{13\pm 1}=a_{21},\nonumber\\
&&a_{22}=W_{N_143\pm 1}+V^{(4)}_{33\pm 1},\nonumber\\
&&b_1=\langle  w_{N_1}\vert E_x\rangle
\langle R_{411}Y_{1\pm 1}\vert M^{(1)}_x\rangle,\nonumber\\
&&b_2=\langle  w_{N_1}\vert E_x\rangle \langle R_{431}Y_{3\pm
1}\vert M^{(3)}_x\rangle,\nonumber
\end{eqnarray}
with solutions
\begin{eqnarray}
c_{N_141\pm 1}&=&\langle  w_{N_1}\vert E_x\rangle\frac{a_{22}
\langle R_{411}Y_{1\pm 1}\vert M^{(1)}_x\rangle-a_{12} \langle
R_{431}Y_{3\pm 1}\vert
M^{(3)}_x\rangle}{\Delta},\nonumber\\
c_{N_143\pm 1}&=&\langle  w_{N_1}\vert E_x\rangle\frac{a_{11}
\langle R_{431}Y_{3\pm 1}\vert M^{(3)}_x\rangle-a_{12} \langle
R_{411}Y_{1\pm 1}\vert M^{(1)}_x\rangle}{\Delta},\\
\Delta&=& a_{11}a_{22}-a_{12}^2.\nonumber
\end{eqnarray}
The third of equations (\ref{cn4}) has the solution in the form
\begin{equation}
c_{N_143\pm 3}=\frac{\langle  w_{N_1}\vert E_x\rangle\langle
R_{433}Y_{3\pm 3}\vert M^{(3)}_x\rangle}{W_{N_143\pm
3}+V^{(4)}_{33\pm 3}}.
\end{equation}
For $n=5$ we obtain quite analogous equations
\begin{eqnarray}\label{cn5}
&&W_{N_151\pm 1}c_{N_151\pm 1}+V^{(5)}_{11\pm 1}c_{N_151\pm
1}+V^{(5)}_{13\pm 1}c_{N_153\pm 1}=\langle  w_{N_1}\vert
E_x\rangle
\langle R_{511}Y_{1\pm 1}\vert M^{(1)}_x\rangle,\nonumber\\
&&W_{N_153\pm 1}c_{N_153\pm 1}+V^{(5)}_{33\pm 1}c_{N_153\pm
1}+V^{(5)}_{31\pm 1}c_{N_151\pm 1}=\langle  w_{N_1}\vert
E_x\rangle
\langle R_{531}Y_{3\pm 1}\vert M^{(3)}_x\rangle,\\
&&W_{N_153\pm 3}c_{N_153\pm 3}+V^{(5)}_{33\pm 3}c_{N_153\pm
3}=\langle  w_{N_1}\vert E_x\rangle \langle R_{533}Y_{3\pm 3}\vert
M^{(3)}_x\rangle\nonumber.
\end{eqnarray}
The third of the above equations yields
\begin{equation}
c_{N_153\pm 3}=\frac{\langle  w_{N_1}\vert E_x\rangle \langle
R_{533}Y_{3\pm 3}\vert M^{(3)}_x\rangle}{W_{N_153\pm
3}+V^{(5)}_{33\pm 3}},
\end{equation}
whereas the first two give the coefficients $c_{N_151\pm
1},c_{N_153\pm 1}$
\begin{eqnarray}
c_{N_15\pm 1}&=&\langle  w_{N_1}\vert E_x\rangle\frac{a_{22}
\langle R_{511}Y_{1\pm 1}\vert M^{(1)}_x\rangle-a_{12} \langle
R_{531}Y_{3\pm 1}\vert
M^{(3)}_x\rangle}{a_{11}a_{22}-a_{12}^2},\nonumber\\
c_{N_153\pm 1}&=&\langle  w_{N_1}\vert E_x\rangle\frac{a_{11}
\langle R_{531}Y_{3\pm 1}\vert M^{(3)}_x\rangle-a_{12} \langle
R_{511}Y_{1\pm 1}\vert M^{(1)}_x\rangle}{a_{11}a_{22}-a_{12}^2},
\end{eqnarray}
where
\begin{eqnarray*}
&&a_{11}=W_{N_151\pm 1}+V^{(5)}_{11\pm 1},\\
&&a_{12}=V^{(5)}_{13\pm 1}=a_{21},\nonumber\\
&&a_{22}=W_{N_153\pm 1}+V^{(5)}_{33\pm 1}.\nonumber
\end{eqnarray*}
In a similar way the higher order coefficients can be determined:
\begin{eqnarray}
c_{N_1n3\pm 3}&=&\frac{\langle  w_{N_1}\vert E_x\rangle \langle
R_{n33}Y_{3\pm 3}\vert M^{(3)}_x\rangle}{W_{N_1n3\pm
3}+V^{(n)}_{33\pm 3}},\nonumber\\
c_{N_1n1\pm 1}&=&\langle  w_{N_1}\vert E_x\rangle\frac{a_{22}
\langle R_{n11}Y_{1\pm 1}\vert M^{(1)}_x\rangle-a_{12} \langle
R_{n31}Y_{3\pm 1}\vert
M^{(3)}_x\rangle}{a_{11}a_{22}-a_{12}^2},\\
c_{N_1n3\pm 1}&=&\langle  w_{N_1}\vert E_x\rangle\frac{a_{11}
\langle R_{n31}Y_{3\pm 1}\vert M^{(3)}_x\rangle-a_{12} \langle
R_{n11}Y_{1\pm 1}\vert M^{(1)}_x\rangle}{a_{11}a_{22}-a_{12}^2},\nonumber\\
a_{11}&=&W_{N_1n1\pm 1}+V^{(n)}_{11\pm 1},\nonumber\\
a_{12}&=&V^{(n)}_{13\pm 1}=a_{21},\nonumber\\
a_{22}&=&W_{N_1n3\pm 1}+V^{(n)}_{33\pm 1},\nonumber\\
\Delta&=&a_{11}a_{22}-a_{12}^2\nonumber
\end{eqnarray}
\begin{eqnarray*}
&&c_{N_1n1\pm 1}=\frac{\langle  w_{N_1}\vert
E_x\rangle}{\Delta}\left\{a_{22}\left[\mp\sqrt{\frac{8\pi}{3}}\frac{M_{10}}{2{\rm
i}}\left(\frac{r_0}{a^*}\right)\frac{\eta_{11}^{5/2}}{a^{*3/2}}\sqrt{\frac{n^2-1}{n^5}}\right]\right.\\
&&\left.-a_{12}\left[\mp
0.015272\cdot\sqrt{\frac{2}{3}\pi}M_{10}\left(\frac{r_0}{a^*}\right)^3\frac{\eta_{31}^{9/2}}{a^{*3/2}}\sqrt{\frac{(n^2-9)(n^2-4)(n^2-1)}{n^9}}\right]\right\}\\
&&=\frac{\langle  w_{N_1}\vert
E_x\rangle}{\Delta}\left(\frac{r_0}{a^*}\right)\sqrt{\frac{2}{3}\pi}M_{10}\frac{1}{a^{*3/2}}\sqrt{\frac{n^2-1}{n^5}}\left\{\pm
a_{22}{\rm i}\eta_{11}^{5/2}\pm a_{12}\cdot
0.015272\cdot\left(\frac{r_0}{a^*}\right)^2\eta_{31}^{9/2}\sqrt{\frac{(n^2-9)(n^2-4)}{n^4}}\right\}\\
&&=\pm\frac{\langle  w_{N_1}\vert E_x\rangle}{\left(W_{N_1n1\pm
1}+V^{(n)}_{111}\right)\left(W_{N_1n3\pm 1}+V^{(n)}_{33
1}\right)-\left(V^{(n)}_{131}\right)^2}\left(\frac{r_0}{a^*}\right)\sqrt{\frac{2}{3}\pi}M_{10}\frac{1}{a^{*3/2}}\sqrt{\frac{n^2-1}{n^5}}\\
&&\times\left\{{\rm i}\eta_{11}^{5/2}\left(W_{N_1n3\pm
1}+V^{(n)}_{331}\right)+0.015272\cdot V^{(n)}_{13
1}\left(\frac{r_0}{a^*}\right)^2\eta_{31}^{9/2}\sqrt{\frac{(n^2-9)(n^2-4)}{n^4}}\right\}
\end{eqnarray*}
\begin{eqnarray*}
&&c_{N_1n3\pm 1}=\langle  w_{N_1}\vert E_x\rangle\frac{a_{11}
\langle R_{n31}Y_{3\pm 1}\vert M^{(3)}_x\rangle-a_{12} \langle
R_{n11}Y_{1\pm 1}\vert M^{(1)}_x\rangle}{a_{11}a_{22}-a_{12}^2}\\
&&=\frac{\langle  w_{N_1}\vert
E_x\rangle}{\Delta}\left\{a_{11}\left[\mp
0.015272\cdot\sqrt{\frac{2}{3}\pi}M_{10}\left(\frac{r_0}{a^*}\right)^3\frac{\eta_{31}^{9/2}}{a^{*3/2}}\sqrt{\frac{(n^2-9)(n^2-4)(n^2-1)}{n^9}}\right]\right.\\
&&\left.-a_{12}\left[\mp\sqrt{\frac{8\pi}{3}}\frac{M_{10}}{2{\rm
i}}\left(\frac{r_0}{a^*}\right)\frac{\eta_{11}^{5/2}}{a^{*3/2}}\sqrt{\frac{n^2-1}{n^5}}\right]\right\}\\
&&=\mp\frac{\langle  w_{N_1}\vert E_x\rangle}{\Delta}\left(\frac{r_0}{a^*}\right)\sqrt{\frac{2}{3}\pi}M_{10}\frac{1}{a^{*3/2}}\sqrt{\frac{n^2-1}{n^5}}\\
&&\left\{a_{11}\cdot
0.015272\cdot\left(\frac{r_0}{a^*}\right)^2\eta_{31}^{9/2}\sqrt{\frac{(n^2-9)(n^2-4)}{n^4}}+a_{12}{\rm
i}\eta_{11}^{5/2}\right\}\\
&&=\mp\frac{\langle  w_{N_1}\vert E_x\rangle}{\left(W_{N_1n1\pm
1}+V^{(n)}_{111}\right)\left(W_{N_1n3\pm 1}+V^{(n)}_{33
1}\right)-\left(V^{(n)}_{131}\right)^2}\left(\frac{r_0}{a^*}\right)\sqrt{\frac{2}{3}\pi}M_{10}\frac{1}{a^{*3/2}}\sqrt{\frac{n^2-1}{n^5}}\\
&&\times\left\{\left[W_{N_1n1\pm 1}+V^{(n)}_{111}\right]\cdot
0.015272\cdot\left(\frac{r_0}{a^*}\right)^2\eta_{31}^{9/2}\sqrt{\frac{(n^2-9)(n^2-4)}{n^4}}\right.\\
&&\left.+V^{(n)}_{131}\cdot{\rm i}\eta_{11}^{5/2}\right\}
\end{eqnarray*}
\section{Determination of the quantities $\langle R_{n11}Y_{1\pm 1}\vert
M^{(1)}_x\rangle, \langle R_{n31}Y_{3\pm 1}\vert M^{(3)}_x\rangle,
\langle R_{n33}Y_{3\pm 3}\vert M^{(3)}_x\rangle$}\label{Appendix
B}

Below we calculate the quantities $\langle R_{n11}Y_{1\pm 1}\vert
M^{(1)}_x\rangle, \langle R_{n31}Y_{3\pm 1}\vert M^{(3)}_x\rangle,
\langle R_{n33}Y_{3\pm 3}\vert M^{(3)}_x\rangle$, which enter in
the above derived formulas for the coefficients $c_{Nn\ell m}$.
Using the definition (\ref{Mx1}, \ref{Mx3}) one obtains

\begin{eqnarray}\langle R_{n11}Y_{1\pm 1}\vert
M_x^{(1)}x\rangle&=&\mp\sqrt{\frac{8\pi}{3}}\frac{M_{10}}{4{\rm
i}r_0^2}\int\limits_0^\infty r^2{\rm
d}r\;\frac{r+r_0}{r^2}e^{-r/r_0}R_{n11}(r).
\end{eqnarray}
Inserting on the r.h.s. the expression for the radial function
$R_{n11}$ (see Eq. (\ref{radialfinala})) we get
\begin{eqnarray}\label{Rn11}
\langle R_{n11}Y_{1\pm 1}\vert M_x^{(1)}x\rangle&=&
\mp\sqrt{\frac{8\pi}{3}}\frac{M_{10}}{4{\rm
i}r_0^2}\int\limits_0^\infty {\rm d}r\;(r+r_0)e^{-\lambda
r}\left(\frac{2\eta_{1 1}}{na^*}\right)^{3/2}\frac{1}{3
!}\sqrt{\frac{(n+1)!}{2n(n-2)!}}\nonumber\\
&&\times\left(\frac{2\eta_{11}r}{na^*}\right)
M\left(-n+2,4,\frac{2\eta_{1 1}r}{na^*}\right),
\end{eqnarray}
where
\begin{eqnarray*}
\lambda&=&\frac{1}{r_0}+\frac{\eta_{11}}{na^*}=\frac{na^*+\eta_{11}r_0}{na^*r_0}.
\end{eqnarray*}
The r.h.s. of Eq. (\ref{Rn11}) consists of two
parts:
\begin{eqnarray}\label{I_1}
&&I_1=\sqrt{\frac{8\pi}{3}}\frac{M_{10}}{4{\rm
i}r_0^2}\left(\frac{2\eta_{11}}{na^*}\right)^{3/2}\left(\frac{2\eta_{11}}{na^*}\right)\frac{1}{3
!}\sqrt{\frac{(n+1)!}{2n(n-2)!}}\int\limits_0^\infty
{\rm d}r\;r^2\,e^{-\lambda r}\nonumber\\
&&\times M\left(-n+2,4,\frac{2\eta_{1 1}r}{na^*}\right)\\
&&=\sqrt{\frac{8\pi}{3}}\frac{M_{10}}{4{\rm
i}r_0^2}\left(\frac{2\eta_{11}}{na^*}\right)^{5/2}\frac{1}{3
!}\sqrt{\frac{(n+1)!}{2n(n-2)!}}\Gamma(3)\lambda^{-3}F\left(-n+2,3,4,\frac{2\eta_{11}r_0}{na^*+\eta_{11}r_0}\right),\nonumber\\
&&\label{I_2}I_2=\sqrt{\frac{8\pi}{3}}\frac{M_{10}}{4{\rm
i}r_0^2}r_0\left(\frac{2\eta_{11}}{na^*}\right)^{5/2}\frac{1}{3
!}\sqrt{\frac{(n+1)!}{2n(n-2)!}}\\
&&\times \int\limits_0^\infty r{\rm d}r\;e^{-\lambda r}
M\left(-n+2,4,\frac{2\eta_{11}}{na^*}\cdot r\right)\nonumber\\
&&=\sqrt{\frac{8\pi}{3}}\frac{M_{10}}{4{\rm
i}r_0^2}r_0\lambda^{-2}\left(\frac{2\eta_{11}}{na^*}\right)^{5/2}\frac{1}{3
!}\sqrt{\frac{(n+1)!}{2n(n-2)!}}F\left(-n+2,2,4,\frac{2\eta_{11}r_0}{na^*+\eta_{11}r_0}\right),\nonumber
\end{eqnarray}
$F(\alpha,\beta,\gamma,z)$ being the hypergeometric series (for
example, \cite{Grad})
\begin{equation}
F=1+\frac{\alpha\beta}{\gamma}\,\frac{z}{1!}+\frac{\alpha(\alpha+1)\beta(\beta+1)}{\gamma(\gamma+1)}\,\frac{z^2}{2!}+\ldots.
\end{equation}
Performing the summation we obtain
\begin{eqnarray*}
&&I_1+I_2=\sqrt{\frac{8\pi}{3}}\frac{M_{10}}{4{\rm
i}r_0^2}\left(\frac{2\eta_{11}}{na^*}\right)^{5/2}\frac{1}{3
!}\sqrt{\frac{(n+1)!}{2n(n-2)!}}2\lambda^{-3}F\left(-n+2,3,4,\frac{2\eta_{11}r_0}{na^*+\eta_{11}r_0}\right)\\
&&+\sqrt{\frac{8\pi}{3}}\frac{M_{10}}{4{\rm
i}r_0^2}r_0\lambda^{-2}\left(\frac{2\eta_{11}}{na^*}\right)^{5/2}\frac{1}{3
!}\sqrt{\frac{(n+1)!}{2n(n-2)!}}F\left(-n+2,2,4,\frac{2\eta_{11}r_0}{na^*+\eta_{11}r_0}\right)\\
&&\approx \sqrt{\frac{8\pi}{3}}\frac{M_{10}}{4{\rm
i}r_0^2}\left(\frac{2\eta_{11}}{na^*}\right)^{5/2}\frac{1}{3
!}\sqrt{\frac{(n+1)!}{2n(n-2)!}}\frac{n^2a^{*2}r_0^3(3na^*+\eta_{11}r_0)}{(na^*+\eta_{11}r_0)^3}\\
&&=\sqrt{\frac{8\pi}{3}}\frac{M_{10}r_0}{8{\rm
i}}\left(\frac{2\eta_{11}}{na^*}\right)^{5/2}\sqrt{\frac{(n+1)!}{2n(n-2)!}},
\end{eqnarray*}
where the assumption $r_0<a^*$ has been used. Finally
\begin{eqnarray}
 &&\langle R_{n11}Y_{1\pm 1}\vert M_x^{(1)}\rangle=\mp\sqrt{\frac{2\pi}{3}}\frac{M_{10}}{{\rm
i}}\left(\frac{r_0}{a^*}\right)\frac{\eta_{11}^{5/2}}{a^{*3/2}}\sqrt{\frac{n^2-1}{n^5}}.
\end{eqnarray}
The following expression will be useful in calculation of
oscillator strengths \begin{eqnarray}
&&\left|I_1+I_2\right|^2=\frac{2}{3}\pi
\left|M_{10}\right|^2\left(\frac{r_0}{a^*}\right)^2\frac{\eta_{11}^5}{a^{*3}}\frac{n^2-1}{n^5}.
\end{eqnarray}
In the next step we calculate the quantity $\langle R_{n31}Y_{3\pm
1}\vert M_x^{(3)}x\rangle$. Using the definitions (\ref{M3}) and
(\ref{radialfinala}) one obtains
\begin{eqnarray*}
&&\langle R_{n31}Y_{3\pm 1}\vert
M_x^{(3)}x\rangle=\mp\sqrt{\frac{3\pi}{7}}\frac{M_{10}}{r_0}\int\limits_0^\infty
{\rm
d}r\;e^{-r/r_0}R_{n31}(r)\\
&&=\mp\sqrt{\frac{3\pi}{7}}\frac{M_{10}}{r_0}\left(\frac{2\eta_{31}}{na^*}\right)^{9/2}\int\limits_0^\infty
{\rm d}r\;r^3\,e^{-\lambda r}\frac{1}{7
!}\sqrt{\frac{(n+3)!}{2n(n-4)!}}\nonumber\\
&&\times M\left(-n+4,8,\frac{2\eta_{31}}{na^*}\cdot r\right)\\
\end{eqnarray*}
with
\begin{eqnarray*}
\lambda=\frac{1}{r_0}+\frac{\eta_{31}}{na^*}=\frac{na^*+\eta_{31}r_0}{na^*r_0}.
\end{eqnarray*}
Performing the integration we arrive at the formulas
\begin{eqnarray*}
&&\langle R_{n31}Y_{3\pm 1}\vert
M_x^{(3)}\rangle=\mp\sqrt{\frac{3\pi}{7}}\frac{M_{10}}{r_0}\left(\frac{2\eta_{31}}{na^*}\right)^{9/2}\sqrt{\frac{(n+3)!}{2n(n-4)!}}\,
\frac{3!}{7!}\lambda^{-4}\,F\left(-n+4,4,8,\frac{2\eta_{31}}{na^*\lambda}\right)\\
&&=\mp\sqrt{\frac{3\pi}{7}}\frac{M_{10}}{r_0}\left(\frac{2\eta_{31}}{na^*}\right)^{9/2}\sqrt{\frac{(n+3)!}{2n(n-4)!}}\,
 \frac{3!}{7!}\frac{n^4a^{*4}r_0^4}{(na^*+\eta_{31}r_0)^4}\,F\left(-n+4,4,8,\frac{2\eta_{31}}{na^*\lambda}\right)
 \\
 &&\approx \mp\;
 0.015272\cdot\sqrt{\frac{2}{3}\pi}M_{10}\left(\frac{r_0}{a^*}\right)^3\frac{\eta_{31}^{9/2}}{a^{*3/2}}\sqrt{\frac{(n^2-9)(n^2-4)(n^2-1)}{n^9}},
 \end{eqnarray*}
 where again the assumption $r_0<a^*$ has been used. The formula
 \begin{eqnarray*}
&&\left|\langle R_{n31}Y_{3\pm 1}\vert
M_x^{(3)}\rangle\right|^2\approx\frac{3\pi}{2\cdot7}\frac{\vert
M_{10}\vert^2}{a^{*3}}\left(\frac{3!}{7!}\right)^2\left(2\eta_{31}\right)^{9}\left(\frac{r_0}{a^*}\right)^6\frac{(n^2-9)(n^2-4)(n^2-1)}{n^9}\\
&&=2.332\cdot 10^{-4}\frac{2}{3}\pi\frac{\vert
M_{10}\vert^2}{a^{*3}}\left(\eta_{31}\right)^{9}\left(\frac{r_0}{a^*}\right)^6\frac{(n^2-9)(n^2-4)(n^2-1)}{n^9}
\end{eqnarray*}
will be used in calculations of the oscillator strengths related
to the F exciton. The remaining formulas, connected to the
$Y_{3,\pm 3}$ harmonics, have the form

\begin{eqnarray}
&&\langle R_{n33}Y_{3\pm 3}\vert
M_x^{(3)}\rangle=\mp\sqrt{\frac{5\pi}{7}}\frac{M_{10}}{r_0}\left(\frac{2\eta_{33}}{na^*}\right)^{9/2}\sqrt{\frac{(n+3)!}{2n(n-4)!}}\,
\frac{3!}{7!}\lambda^{-4}\,F\left(-n+4,4,8,\frac{2\eta_{33}}{na^*\lambda}\right)\nonumber\\
&&\left|\langle R_{n33}Y_{3\pm 3}\vert
M_x^{(3)}\rangle\right|^2\approx\frac{5\pi}{2\cdot7}\frac{\vert
M_{10}\vert^2}{a^{*3}}\left(\frac{3!}{7!}\right)^2\left(2\eta_{33}\right)^{9}\left(\frac{r_0}{a^*}\right)^6\frac{(n^2-9)(n^2-4)(n^2-1)}{n^9}\\
&&=3.887\cdot 10^{-4}\frac{2}{3}\pi\frac{\vert
M_{10}\vert^2}{a^{*3}}\left(\eta_{33}\right)^{9}\left(\frac{r_0}{a^*}\right)^6\frac{(n^2-9)(n^2-4)(n^2-1)}{n^9}.\nonumber
\end{eqnarray}

\section{Derivation of the matrix elements
$V^{(nn_1)}_{\ell\ell_1m m_1}$}\label{Appendix C} In order to
calculate the matrix elements $V^{(nn_1)}_{\ell\ell_1m m_1}$, as
defined in Eq. (\ref{definitionVn}), we start with the integral
containing the angular dependence
\begin{eqnarray}\label{Iell}
I_{\ell\ell_1mm_1}&=&\langle Y_{\ell_1m_1}\vert\sin^2\theta\vert
Y_{\ell m}\rangle=\int {\rm
d}\Omega\;Y^*_{\ell_1m_1}(1-\cos^2\theta)Y_{\ell
m}\nonumber\\
&=&\delta_{\ell\ell_1}\delta_{mm_1}-\int{\rm
d}\Omega\;Y_{\ell_1m_1}\cos^2\theta\,Y_{\ell m}.
\end{eqnarray}
Making use of the definition of the spherical harmonic functions
\begin{equation}
 Y_{\ell
m}(\theta,\phi)=\sqrt{\frac{(2\ell+1)(\ell-m)!}{4\pi(\ell+m)!}}P^m_\ell(\cos\theta)e^{{\rm
i}m\phi},
\end{equation}
in terms of the associated Legendre polynomials $P^m_\ell$, the
second of the integrals on the r.h.s. of Eq. (\ref{Iell}) can be
put into the form
\begin{eqnarray}\label{II}
I&=&\int{\rm d}\Omega\;Y_{\ell_1m_1}^*\cos^2\theta\,Y_{\ell m}\\
&&=\delta_{mm_1}2\pi\sqrt{\frac{(2\ell+1)(\ell-m)!}{4\pi(\ell+m)!}\frac{(2\ell_1+1)(\ell_1-m_1)!}{4\pi(\ell_1+m_1)!}}\int\limits_{-1}^{+1}{\rm
d}x\,  P^m_\ell(x)x^2P^{m_1}_{\ell_1}(x).\nonumber
\end{eqnarray}
Making use of the recurrence relation (\cite{Grad})
\begin{eqnarray}
&&xP^m_\ell(x)=\frac{1}{2\ell+1}\left[(\ell-m+1)P^m_{\ell+1}(x)+(\ell+m)P^m_{\ell-1}(x)\right],
\end{eqnarray}
we arrive at the integral
\begin{eqnarray}
&&\int\limits_{-1}^{+1}{\rm d}x\,
P^m_\ell(x)x^2P^{m_1}_{\ell_1}(x)=\int\limits_{-1}^{+1}{\rm
d}x\biggl\{\frac{1}{2\ell+1}\left[(\ell-m+1)P^m_{\ell+1}(x)+(\ell+m)P^m_{\ell-1}(x)\right]\nonumber\\
&&\times\;\frac{1}{2\ell_1+1}\left[(\ell_1-m+1)P^m_{\ell_1+1}(x)+(\ell_1+m)P^m_{\ell_1-1}(x)\right]\biggr\}.
\end{eqnarray}
Performing the multiplication and integration, using the
orthogonality relation $$
\int\limits_{-1}^{+1}P^m_\ell(x)P^m_{\ell_1}(x){\rm
d}x=\frac{2}{2\ell+1}\frac{(\ell+m)!}{(\ell-m)!}\delta_{\ell\ell_1},$$
one obtains

\begin{eqnarray}
&&\int\limits_{-1}^{+1}{\rm d}x\,
P^m_\ell(x)x^2P^{m_1}_{\ell_1}(x)=\frac{2(2\ell^2+2\ell-1-2m^2)(\ell+m)!}{(2\ell-1)(2\ell+1)(2\ell+3)(\ell-m)!}\delta_{\ell\ell_1}\nonumber\\
&&+\frac{2(\ell-m+1)(\ell+m+2)(\ell+m+1)!}{(2\ell+1)(2\ell+3)(2\ell+5)(\ell+1-m)!}\delta_{\ell+1,\ell_1-1}\\
&&+\frac{2(\ell-m-1)(\ell+m)(\ell+m-1)!}{(2\ell-3)(2\ell-1)(2\ell+1)(\ell-1-m)!}\delta_{\ell_1+1,\ell-1}.\nonumber
\end{eqnarray}
 The above result inserting into the Eq. (\ref{II}) gives for the
 angular integration

\begin{eqnarray}
I&=&
\delta_{mm_1}2\pi\sqrt{\frac{(2\ell+1)(\ell-m)!}{4\pi(\ell+m)!}\frac{(2\ell_1+1)(\ell_1-m_1)!}{4\pi(\ell_1+m_1)!}}\nonumber\\
&&\times\;\biggl[\frac{2(2\ell^2+2\ell-1-2m^2)(\ell+m)!}{(2\ell-1)(2\ell+1)(2\ell+3)(\ell-m)!}\delta_{\ell\ell_1}\nonumber\\
&&+\frac{2(\ell-m+1)(\ell+m+2)(\ell+m+1)!}{(2\ell+1)(2\ell+3)(2\ell+5)(\ell+1-m)!}\delta_{\ell+1,\ell_1-1}\\
&&+\frac{2(\ell-m-1)(\ell+m)(\ell+m-1)!}{(2\ell-3)(2\ell-1)(2\ell+1)(\ell-1-m)!}\delta_{\ell_1+1,\ell-1}\biggr].\nonumber
\end{eqnarray}
Using the above results we obtain the diagonal matrix element
$V^{nn_1}_{\ell\ell_1mm_1}$ when $n=n_1$
\begin{eqnarray}
V^{n}_{\ell_1\ell_1m_1m_1}&=&\frac{R^*}{4}\gamma^2\delta_{\ell\ell_1mm_1}\left(1-\frac{2\ell_1^2+2\ell_1-1-2m_1^2}{(2\ell_1-1)(2\ell_1+3)}\right)\int\limits_0^\infty{\rm
d}\rho
\rho^4\,R_{n\ell}R_{n_1\ell_1}\\
&=&\delta_{\ell\ell_1mm_1}\frac{R^*}{4}\gamma^2\frac{2(\ell^2+\ell+m^2-1)}{(2\ell-1)(2\ell+3)}\int\limits_0^\infty{\rm
d}\rho\, \rho^4\,\left[R_{n\ell}(\rho)\right]^2\nonumber\\
&=&\delta_{\ell\ell_1mm_1}\frac{R^*}{4}\gamma^2\frac{(\ell^2+\ell+m^2-1)}{(2\ell-1)(2\ell+3)}\left(\frac{n}{\eta_{\ell
m}}\right)^2[5n^2+1-3\ell(\ell+1)]\nonumber
\end{eqnarray}
where we used the formula (for example \cite{Messiah} where
$\eta_{\ell m}=1$)
\begin{equation}
\langle \rho^2\rangle=\frac{1}{2}[5n^2+1-3\ell(\ell+1)].
\end{equation}
In a similar way the the off-diagonal elements can be computed
with the results displayed in Eq. (\ref{basic_equations}). The
integrals containing the radial eigenfunctions $R_{n\ell m}$ can
be expressed by the integrals over Laguerre polynomials
\begin{eqnarray*}
&&I_{n\ell s m}=\int\limits_0^\infty{\rm d}\rho \rho^4\,R_{n\ell
m}R_{ns m}=\int\limits_0^\infty{\rm d}\rho
\rho^4\left(\frac{2\eta_{\ell
m}}{n}\right)^{3/2}\sqrt{\frac{(n-\ell-1)!}{2n(n+\ell)!}}\left(\frac{2\eta_{\ell
m}\rho}{n}\right)^\ell\nonumber\\
&&\times L_{n-\ell-1}^{2\ell+1}\left(\frac{2\eta_{\ell
m}\rho}{n}\right) e^{-\eta_{\ell m}\rho/n}\\
&&\times\left(\frac{2\eta_{\ell
m}}{n}\right)^{3/2}\sqrt{\frac{(n-s-1)!}{2n(n+s)!}}\left(\frac{2\eta_{\ell
m}\rho}{n}\right)^s\, L_{n-s-1}^{2s+1}\left(\frac{2\eta_{\ell
m}\rho}{n}\right) e^{-\eta_{\ell m}\rho/n}\\
&&=\left(\frac{n}{2\eta_{\ell
m}}\right)^2\sqrt{\frac{(n-\ell-1)!}{2n(n+\ell)!}}\sqrt{\frac{(n-s-1)!}{2n(n+s)!}}\int\limits_0^\infty
x^{\ell+s+4} L_{n-\ell-1}^{2\ell+1}(x)
L_{n-s-1}^{2s+1}(x)e^{-x}{\rm d}x
\end{eqnarray*}
For example, taking $n=4, \ell=1, s=3, m=1$, one has
\begin{eqnarray*}
I_{41 3 1}&=&\int\limits_0^\infty{\rm d}\rho \rho^4\,R_{41 1}R_{43
1}=\left(\frac{4}{2\eta_{1 0}}\right)^2\sqrt{\frac{2!}{2\cdot
4\cdot 5!}}\sqrt{\frac{0!}{2\cdot 4\cdot7!}}\int\limits_0^\infty
x^8 L_{2}^{3}(x) L_{0}^{7}(x)e^{-x}{\rm d}x\\
&&=\left(\frac{2}{\eta_{1 0}}\right)^2\sqrt{\frac{1}{ 4\cdot
5!}}\sqrt{\frac{1}{8!}}\int\limits_0^\infty x^8
\frac{1}{2}\left[x^2-10x+20\right] e^{-x}{\rm
d}x\\&&=\left(\frac{1}{\eta_{1
0}}\right)^2\,8!\sqrt{\frac{1}{5!8!}}\left(90-90+20\right)=20\left(\frac{1}{\eta_{1
0}}\right)^2\sqrt{\frac{8!}{5!}}\approx 367\left(\frac{1}{\eta_{1
0}}\right)^2.
\end{eqnarray*}

\section{Calculation of the coefficients $\langle w_N\vert E\rangle$}\label{Appendix D}
The homogeneous solution of the field equation (\ref{Ehom}) can be
put into the form
\begin{equation}
E_{\rm hom}(Z)=Ae^{{\rm i}k_bZ}+Be^{-{\rm i}k_bZ},
\end{equation}
\noindent where
\begin{eqnarray}
A&=&\frac{2k_0}{(k_0+k_b)W}e^{-{\rm i}k_bL}E_{\rm in},\nonumber\\
B&=&\frac{2k_0(k_b-k_0)}{(k_0+k_b)^2W}e^{{\rm i}k_bL}E_{\rm in}.
\end{eqnarray}
\noindent With these expressions one obtains for $\langle w_N\vert
E\rangle$
\begin{equation}\label{wuen}
\langle w_N\vert E\rangle=AI+BI^*,
\end{equation}
\noindent with the notation
\begin{equation}
I=\langle w_N\vert \,e^{{\rm
i}k_bZ}\rangle=\sqrt{\frac{2}{L}}\int\limits_0^L\sin\frac{N\pi}{L}Z\,e^{{\rm
i}k_bZ}\,{\rm d}Z.
\end{equation}
\noindent With the use of the relations
\begin{equation}
\sin x=\frac{e^{{\rm i}x}-e^{-{\rm i}x}}{2{\rm i}},\quad
e^{\pm{\rm i}N\pi}=\cos N\pi,
\end{equation}
\noindent the integral $I$ becomes

\begin{equation}\label{ien}
I=I_N=\sqrt{2L}\left(1-\cos\,N\pi\exp\left[{\rm
i}k_b{L}\right]\right)\frac{N\pi}{(N\pi)^2-(k_bL)^2}.
\end{equation}
The quantity $I_N^2/L$ which appears in the expressions for the
optical functions, has the form
\begin{equation}\label{INsquare}
\frac{I_N^2}{L}=\frac{2N^2\pi^2}{\left[\left(k_bL\right)^2-N^2\pi^2\right]^2}\left(1-\cos\,N\pi\exp\left[{\rm
i}k_b{L}\right]\right)^2.
\end{equation}
The expression $k_bL$ can be transformed in the following way:
\begin{eqnarray}
k_bL&=&\frac{\hbar\omega}{\hbar
c}\sqrt{\epsilon_b}L=\frac{E}{E_L},\nonumber\\
E_L&=&\frac{\hbar c}{\sqrt{\epsilon_b}L}=\frac{0.66\cdot
10^{-15}\hbox{eV s}\cdot 3\cdot
10^8\hbox{m/s}}{\sqrt{\epsilon_b}L}\nonumber\\
&=&\frac{1.98\cdot 10^{2}\hbox{meV}\cdot
\mu\hbox{m}}{\sqrt{\epsilon_b}L(\mu\hbox{m})}.
\end{eqnarray}
For the Cu$_2$O data $\epsilon_b=7.5$ and
\begin{equation}
E_L=\frac{72.3}{L}\,\hbox{meV},
\end{equation}
where $L$ is expressed in $\mu\hbox{m}$. For $L=30\,\mu\hbox{m}$
we have $E_L=2.41\,\hbox{meV}$. Since $I_N=I_N^*$, it follows from
(\ref{wuen})
\begin{equation}
\langle w_N\vert E_{\rm hom}(Z)\rangle=I_N(A+B)=I_NE_{\rm hom}(0),
\end{equation}
\noindent where
\begin{eqnarray}\label{Ejednzero}
E_{\rm hom}(0)&=&\frac{2k_0}{(k_0+k_b)W}E_{\rm in}\left(e^{-{\rm i}k_bL}+\frac{k_b-k_0}{k_b+k_0}e^{{\rm i}k_bL}\right)\\
&=&E_{\rm in}\frac{1+r_{\infty}}{1-r_{\infty}^2\exp({\rm
i}\Theta)}\left(1-r_{\infty}e^{{\rm i}\Theta}\right).\nonumber
\end{eqnarray}

\section{Optical functions}\label{Appendix E}

 \noindent Below we derive the formulas (\ref{reflectionR}) and (\ref{transmission}). They will be obtained from Eq. (\ref{RTdefinitions})
 by using the total electric field $E(Z)=E_{\rm hom}(Z)+E_{\rm nhom}(Z)$. In particular, for the reflection
 coefficient one obtains
\begin{equation}\label{definitionR}
R=\vert r\vert^2,
\end{equation}
\noindent with

\begin{eqnarray}
r&=&r_0+r_{\rm exc},\nonumber\\
r_0&=&\frac{r_\infty\left(1-e^{{\rm i}\Theta}\right)}{1-r_\infty^2e^{{\rm i}\Theta}},\\
r_{\rm exc}&=&\frac{k_0^2}{\epsilon_0E_{\rm
in}}\int\limits_0^L\,G^E(0,Z)P_{\rm exc}(Z){\rm d}Z.\nonumber
\end{eqnarray}
\noindent Since
\begin{equation}
G^E(0,Z)=\frac{{\rm
i}(1-r_\infty)}{2k_b\left(1-r_{\infty}^2\,e^{{\rm
i}\Theta}\right)}\left(e^{{\rm i}k_bZ}-r_\infty\,e^{{\rm
i}\Theta-{\rm i}k_bZ}\right),
\end{equation}
\noindent we obtain
\begin{eqnarray}
&&r_{\rm exc}=\frac{k_0^2}{\epsilon_0E_{\rm in}}\int\limits_0^L\,G^E(0,Z)P_{\rm eks}(Z){\rm d}Z\nonumber\\
&&=\frac{k_0^2}{\epsilon_0E_{\rm in}}\frac{{\rm
i}(1-r_\infty)}{2k_b\left(1-r_{\infty}^2\,e^{{\rm
i}\Theta}\right)}
\sum\limits_N\int\limits_0^L\,\left(e^{{\rm i}k_bZ}-r_\infty\,e^{{\rm i}\Theta-{\rm i}k_bZ}\right)P_Nw_N(Z){\rm d}Z\\
&&=\frac{k_0^2}{2\epsilon_0E_{\rm in}}\frac{{\rm
i}(1-r_\infty)}{k_b\left(1-r_{\infty}^2\,e^{{\rm
i}\Theta}\right)}\sum\limits_N\left[\langle w_N\vert e^{{\rm
i}k_bZ}\rangle-
r_\infty\,e^{{\rm i}\Theta}\langle w_N\vert e^{-{\rm i}k_bZ}\rangle\right]P_N\nonumber\\
&&=\frac{k_0^2}{2\epsilon_0E_{\rm in}}\frac{{\rm
i}(1-r_\infty)}{k_b\left(1-r_{\infty}^2\,e^{{\rm
i}\Theta}\right)}\left(1-r_\infty\,e^{{\rm
i}\Theta}\right)\sum\limits_NI_NP_N,\nonumber
\end{eqnarray}
\noindent from which, with respect to Eqn. (\ref{Polaryzacja_1}),
 we get
\begin{eqnarray}
r_{\rm exc}&=&\frac{k_0^2E_{\rm hom}(0)}{2E_{\rm in}}\frac{{\rm
i}(1-r_\infty)}{k_b\left(1-r_{\infty}^2\,e^{{\rm
i}\Theta}\right)}\left(1-r_\infty\,e^{{\rm
i}\Theta}\right)\epsilon_b\sum\limits_NI_N^2\sum\limits_{n\ell
m}\Delta_{LT}^{(P)}\chi_{N n\ell m}.
\end{eqnarray}
\noindent Inserting in the above equation the expression
(\ref{Ejednzero}), we obtain
\begin{eqnarray}\label{reks1}
r_{{\rm exc}}&=&\frac{\epsilon_bk_0^2} {2k_b}\frac{{\rm
i}(1-r^2_\infty)}{\left(1-r_{\infty}^2\,e^{{\rm
i}\Theta}\right)^2} \left(1-r_\infty\,e^{{\rm
i}\Theta}\right)^2\sum\limits_N{I_N^2}\sum\limits_{n\ell
m}\Delta_{LT}^{(P)}\chi_{N n\ell m}\nonumber\\
&&=\frac{\rm
i}{2}\,\frac{(k_bL)(1-r^2_\infty)}{\left(1-r_{\infty}^2\,e^{{\rm
i}\Theta}\right)^2} \left(1-r_\infty\,e^{{\rm
i}\Theta}\right)^2\sum\limits_N\left(\frac{I_N^2}{L}\right)\sum\limits_{n\ell
m}\Delta_{LT}^{(P)}\chi_{N n\ell m}.
\end{eqnarray}
Now the total complex reflection coefficient $r$ obtains the form
\begin{eqnarray}
r&=&r_0+r_{\rm exc}=r_0\Biggl[1+{\rm
i}\left(\frac{E}{2E_L}\right)\frac{(1-r_\infty^2)(1-r_\infty
e^{{\rm i}\Theta})^2(1-r_\infty^2e^{{\rm
i}\Theta})}{r_\infty(1-e^{{\rm
i}\Theta})}\nonumber\\
&&\times\sum\limits_N\left(\frac{I_N^2}{L}\right)\sum\limits_{n\ell
m}\Delta_{LT}^{(P)}\chi_{N n\ell m}\Biggr],
\end{eqnarray}
which, using the Eq. (\ref{definitionR}) immediately gives the
result (\ref{reflectionR}).

 Analogically, we determine the transmissivity
\begin{equation}
T=\left|\frac{E(L)}{E_{\rm in}}\right|^2,
\end{equation}
\noindent resulting from the equation
\begin{equation}\label{definitionT}
 T=\vert t\vert^2,
\end{equation}
\noindent where

\begin{eqnarray}
t&=&t_0+t_{\rm exc},\nonumber\\
t_0&=&\frac{1-r_\infty^2}{1-r_\infty^2e^{{\rm i}\Theta}}e^{{\rm i}\Theta/2},\\
t_{\rm exc}&=&\frac{k_0^2}{\epsilon_0E_{\rm
in}}\int\limits_0^L\,G^E(L,Z)P_{\rm exc}(Z){\rm d}Z.\nonumber
\end{eqnarray}
\noindent Since
\begin{eqnarray}
&&G^E(L,Z)=\frac{\rm i}{2k_bW}\left(e^{-{\rm i}k_bZ}+\frac{k_b-k_0}{k_b+k_0}e^{{\rm i}k_bZ}\right)\nonumber\\
&&\times\>\left(\frac{k_b-k_0}{k_b+k_0}e^{{\rm i}k_bL-{\rm i}k_bL}+e^{-{\rm i}k_bL+{\rm i}k_bL}\right)\\
&&=\frac{{\rm i}e^{{\rm
i}\Theta/2}\left(1-r_\infty\right)}{2k_b\left(1-r_\infty^2 e^{{\rm
i}\Theta}\right)}\left(e^{-{\rm i}k_bZ}-r_\infty e^{{\rm
i}k_bZ}\right),\nonumber
\end{eqnarray}
\noindent we have
\begin{eqnarray}
&&t_{{\rm exc}}=\frac{k_0^2}{E_{\rm in}}\int\limits_0^L\,G^E(L,Z)P_{\rm exc}(Z){\rm d}Z\nonumber\\
&&=\frac{(k_bL)}{2}\frac{{\rm i}e^{{\rm
i}\Theta/2}\left(1-r_\infty\right)}{\left(1-r_\infty^2 e^{{\rm
i}\Theta}\right)^2}\left(1+r_\infty\right)\left(1-r_\infty\,e^{{\rm
i}\Theta}\right)\sum\limits_N\left(\frac{I_N^2}{L}\right)\sum\limits_{n\ell
m}\Delta_{LT}^{(P)}\chi_{N n\ell m},\nonumber
\end{eqnarray}
and
\begin{eqnarray}
&&t=t_0+t_{\rm
exc}=t_0\left[1+\frac{1}{t_0}\frac{(k_bL)}{2}\frac{{\rm i}e^{{\rm
i}\Theta/2}\left(1-r_\infty\right)}{\left(1-r_\infty^2 e^{{\rm
i}\Theta}\right)^2}\left(1+r_\infty\right)\left(1-r_\infty\,e^{{\rm
i}\Theta}\right)\sum\limits_N\left(\frac{I_N^2}{L}\right)\sum\limits_{n\ell
m}\Delta_{LT}^{(P)}\chi_{N n\ell m}\right]\nonumber\\
&&=t_0\left[1+{\rm i}\left(\frac{E}{2E_L}\right)\frac{(1-r_\infty
e^{{\rm i}\Theta})}{(1-r_\infty^2e^{{\rm
i}\Theta})}\sum\limits_N\left(\frac{I_N^2}{L}\right)\sum\limits_{n\ell
m}\Delta_{LT}^{(P)}\chi_{N n\ell m}\right].
\end{eqnarray}
Inserting the above result into Eq. (\ref{definitionT}), we obtain
the transmissivity (\ref{transmission}).

\section{Derivation of the quantities $\chi_{Nn\ell m}$}\label{Appendix F}
Using the dispersion relation
\begin{eqnarray}
&&\frac{c^2k^2}{\omega^2}=\epsilon_b+\frac{2}{\epsilon_0}\int {\rm
d}^3r M^* Y,\nonumber\\
&&\frac{2}{\epsilon_0}\int {\rm d}^3r M^* Y=\frac{2}{\epsilon_0}\int {\rm d}^3r M_x^{(1)*}\sum_{n\ell m}c_{Nn\ell m}R_{n\ell m}Y_{\ell m}\\
&&+\frac{2}{\epsilon_0}\int {\rm d}^3r M_x^{(3)*}\sum_{n\ell
m}c_{Nn\ell m}R_{n\ell m}Y_{\ell m}\nonumber
\end{eqnarray}
we define the quantities $\chi_{Nn\ell m}$ and
$\tilde{\chi}_{Nn\ell m}$ :
\begin{eqnarray}
\frac{2}{\epsilon_0}\int {\rm d}^3r M^{(1)*}_x c_{N_1n1\pm
1}R_{n11}Y_{1\pm1}&=&\epsilon_b\Delta_{LT}^{(P)}\tilde{\chi}_{N_1n1\pm
1}\langle E_x\vert
w_{N_1}\rangle\nonumber\\
\frac{2}{\epsilon_0}\int {\rm d}^3r M^{(3)*}_x c_{N_1n3\pm
1}R_{n31}Y_{3\pm1}&=&\epsilon_b\Delta_{LT}^{(P)}\tilde{\chi}_{N_1n3\pm
1}\langle E_x\vert
w_{N_1}\rangle\\
\frac{2}{\epsilon_0}\int {\rm d}^3r M^{(3)*}_x c_{N_1n3\pm
3}R_{n33}Y_{3\pm3}&=&\epsilon_b\Delta_{LT}^{(P)}\tilde{\chi}_{N_1n3\pm
3}\langle E_x\vert w_{N_1}\rangle.\nonumber
\end{eqnarray}
The notation $\tilde{\chi}$ denotes that the formula contains
contributions from both excitons P and F. Using the formulas
(\ref{Mx1},\ref{Mx3}), and the expressions $\langle R_{n11}Y_{1\pm
1}\vert M^{(1)}_x\rangle, \langle R_{n31}Y_{3\pm 1}\vert
M^{(3)}_x\rangle, \langle R_{n33}Y_{3\pm 3}\vert M^{(3)}_x\rangle$
derived in Appendix \ref{Appendix B}, we obtain obtaining the
following formulas: for n=2,3
\begin{eqnarray}\label{chin23}
\chi_{Nn1\pm
1}&=&\frac{f_{n11}}{W_{n1\pm 1}+V^{(n)}_{11\pm 1}}\nonumber\\
f_{n11}&=&\frac{16}{3}\frac{n^2-1}{n^5}.
\end{eqnarray}
For $n\geq 4$ we observe the overlapping of the P and F excitons,
which is reflected in the formulas
\begin{eqnarray}\label{chingeq4} \chi_{Nn3\pm 3}&=&\frac{f_{n33}}{W_{Nn3\pm
3}+V^{(n)}_{33\pm
3}},\nonumber\\
 f_{n33}&=&2.073\cdot
10^{-3}\;\frac{\eta_{33}^9}{\eta_{11}^5}\left(\frac{r_0}{a^*}\right)^4\frac{(n^2-9)(n^2-4)(n^2-1)}{n^9},\nonumber\\
\tilde{\chi}_{Nn1\pm
1}&=&\frac{16}{3}\frac{n^2-1}{n^5}\frac{1}{\left(W_{Nn1\pm
1}+V^{(n)}_{111}\right)\left(W_{Nn3\pm 1}+V^{(n)}_{33
1}\right)-\left(V^{(n)}_{131}\right)^2}\nonumber\\
&&\times\,\left\{\left(W_{Nn3\pm 1}+V^{(n)}_{331}\right)-{\rm
i}\cdot\,0.015272\cdot V^{(n)}_{13
1}\left(\frac{r_0}{a^*}\right)^2\frac{\eta_{31}^{9/2}}{\eta_{11}^{5/2}}\sqrt{\frac{(n^2-9)(n^2-4)}{n^4}}\right\},\nonumber\\
\tilde{\chi}_{Nn3\pm
1}&=&0.015272\frac{16}{3}\left(\frac{r_0}{a^*}\right)^2\frac{\eta_{31}^{9/2}}{\eta_{11}^5}
\frac{n^2-1}{n^5}\sqrt{\frac{(n^2-9)(n^2-4)}{n^4}}\\
&&\times\frac{1}{\left(W_{Nn1\pm
1}+V^{(n)}_{111}\right)\left(W_{Nn3\pm 1}+V^{(n)}_{33
1}\right)-\left(V^{(n)}_{131}\right)^2}\nonumber\\
&&\times\Biggl\{\left[W_{Nn1\pm 1}+V^{(n)}_{111}\right]\cdot
0.015272\cdot\left(\frac{r_0}{a^*}\right)^2\eta_{31}^{9/2}\sqrt{\frac{(n^2-9)(n^2-4)}{n^4}}+{\rm
i}\cdot\,V^{(n)}_{131}\cdot\eta_{11}^{5/2}\Biggr\}.\nonumber
\end{eqnarray}

 \end{document}